\documentclass[
reprint,
 amsmath,amssymb,
 aps,
]{revtex4-2}

\usepackage{longtable}
\usepackage{graphicx}
\usepackage{dcolumn}
\usepackage{bm}

\begin{document}

\preprint{APS/123-QED}

\title{Effect of Earth-Moon's gravity on TianQin's range acceleration noise. III. \\ An analytical model}


\author{Lei Jiao}
\author{Xuefeng Zhang}
 \email{zhangxf38@sysu.edu.cn}
\affiliation{
 MOE Key Laboratory of TianQin Mission, TianQin Research Center for Gravitational Physics \& School of Physics and Astronomy, Frontiers Science Center for TianQin, Gravitational Wave Research Center of CNSA, Sun Yat-sen University (Zhuhai Campus), Zhuhai 519082, China.
}


\date{\today}

\begin{abstract}
TianQin is a proposed space-based gravitational wave detector designed to operate in circular high Earth orbits. As a sequel to [Zhang et al. Phys. Rev. D 103, 062001 (2021)], this work provides an analytical model to account for the perturbing effect of the Earth's gravity field on the range acceleration noise between two TianQin satellites. For such an ``orbital noise,'' the Earth's contribution dominates above $5\times 10^{-5}$ Hz in the frequency spectrum, and the noise calibration and mitigation, if needed, can benefit from in-depth noise modeling. Our model derivation is based on Kaula's theory of satellite gravimetry with Fourier-style decomposition, and uses circular reference orbits as an approximation. To validate the model, we compare the analytical and numerical results in two main scenarios. First, in the case of the Earth's static gravity field, both noise spectra are shown to agree well with each other at various orbital inclinations and radii, confirming our previous numerical work while providing more insight. Second, the model is extended to incorporate the Earth's time-variable gravity. Particularly relevant to TianQin, we augment the formulas to capture the disturbance from the Earth's free oscillations triggered by earthquakes, of which the mode frequencies enter TianQin's measurement band above 0.1 mHz. The analytical model may find applications in gravity environment monitoring and noise-reduction pipelines for TianQin. 
\end{abstract}

\maketitle



\section{Introduction}

TianQin plans to launch three identical satellites equipped with drag-free control of high precision, which orbit the Earth at an altitude of $\sim 10^5\mathrm{km}$. They form a nearly equilateral triangle constellation facing the white dwarf binary RX J0806.3+1527 and detect gravitational waves (GWs) through measuring distance change between satellites using laser interferometry of picometer-level precision \cite{Luo2016}. In addition to GWs, the gravity field of the Earth-Moon's system also causes distance change between the satellites. The effects are two-fold. First, it leads to the constellation deviating from the nominal equilateral triangle, which must be reduced by orbit optimization and control \cite{Ye2019,Tan2020,Ye2021}. Second, it may induce in-band range variations mixing with GW signals, and hence pose a potential noise source to TianQin \cite{Zhang2021}. 

In order to assess the impact of the Earth-Moon's gravity, we have developed a program TQPOP (TianQin Quadruple Precision Orbit Propagator) in the previous work \cite{Zhang2021}. The simulator takes in optimized initial orbit parameters, and uses detailed force models to propagate the satellites' pure-gravity orbits with low numerical noise that is not available from commonly used double precision. The resulting ephemerides were converted to the inter-satellite range acceleration $\ddot\rho$, and we computed the amplitude spectral density (ASD) and made the comparison with the acceleration noise requirement of TianQin. According to the estimates, we expect the effect of the Earth-Moon's gravity to be below the noise requirement in the sensitive frequency band of TianQin ($10^{-4}-1$ Hz) and hence not constituting a showstopper to the mission \cite{Zhang2021}. Furthermore, we also explored how the Earth-Moon's gravity disturbance varies with different orbital radii and orientations for TianQin \cite{Luo2022}. A general trend is that the disturbance shifts to lower frequencies as the orbital radius increases, which may set the lower bound of the measurement band. Some useful guidelines were drawn for the orbit and constellation design for future geocentric missions like TianQin. All these previous works rely heavily on numerical calculations, and it would be desirable to also examine the relevant issues by analytical modeling so as to, e.g., gain more understanding. 

In the line of analytical work, our team have studied the effect of the Sun's and the Moon's point masses on TianQin's constellation stability \cite{Ye2022}. By solving Lagrange's perturbation equations, the explicit orbit solutions have been derived and linked to the arm-length and breathing-angle variations, which can account for the long-term lunisolar influence with sufficient accuracy. In the range-acceleration ASD, the Moon's point mass dominates below $5\times10^{-5}$ Hz \cite{Zhang2021}. Hence in this work, we will instead focus on modeling the disturbance of the Earth's gravity, as it dominates above $5\times10^{-5}$ Hz in the frequency spectrum, and is more relevant to TianQin. 

Important lessons can be learned from low-low satellite-to-satellite tracking gravimetry missions represented by GRACE (Gravity Recovery And Climate Experiment) and GRACE Follow-On \cite{Tapley2004,Kornfeld2019}. The missions mainly consist of two identical satellites with separation of about $220$ km and flying in the same circular polar orbit at an altitude of $300-500$ km. The formation measures inter-satellite distance variation using microwave/laser ranging systems. From the viewpoints of experiment design and data processing, TianQin shares many common aspects with GRACE and GRACE Follow-On, and therefore (semi-)analytical work in space gravimetry can offer valuable reference to TianQin's modeling. 

For instance, Kaula's linear perturbation approach \cite{Kaula1961,Kaula1966,Rosborough1986,Rosborough1987,Sharma1995} can be used for analytically calculating the Earth's gravity field coefficients from satellite observables. An important procedure therein is by expressing the Earth's gravity field experienced by a satellite as a function of the orbital elements \cite{Kaula1961,Kaula1966}. Also based on Kaula's representation of geo-potential, an analytical method for the Earth's gravity-field recovery was provided by \cite{Sharifi2006} (see also \cite{Colombo1984,Cheng2002,Wagner2006,Schrama1990,Schrama1991,Schrama1989,Visser1994,Visser2001,Visser2005,Visser1992}). The method shares the same observable with the inter-satellite line-of-sight differential gravitational acceleration approach \cite{Keller2005,Hajela1974,Rummel1980,Garcia2002,Han2003}, and contains an analytical expression of the differential gravitational acceleration of two GRACE satellites which fly along a circular reference orbit encircling the Earth in uniform rotation. 

Based on the above works, we will construct the analytical model for the influence of the Earth's static gravity field and consider TianQin's special case of a high altitude and long measurement baseline. Then we will, taking the Earth's free oscillations as an example, extend the model to include the Earth's time-variable gravity field. In this latter regard, the reference \cite{Ghobadi2019} has discussed the influence of the Earth's free oscillations triggered by large earthquakes on the inter-satellite measurement of GRACE. We will follow a similar treatment in this work. 

This paper is organized as follows. Section \ref{sec:model} shows the detailed derivation of the analytical model. Section \ref{sec:verification1} tests the model with numerical results obtained from TQPOP. Subsection \ref{subsec:one group} presents the result for TianQin, and Subsection \ref{subsec:several groups} presents the results for other orbital inclinations and radii. Section \ref{sec:expansion} extends the model to include time-variable gravity field from the Earth's free oscillations. Section \ref{sec:verification2} carries out the corresponding numerical verification. At last, Section \ref{sec:conclusions} draws the conclusions with outlooks. 


\section{Model Derivation: Static Gravity}\label{sec:model}

In this section, we detail the derivation of an explicit expression for the range acceleration $\ddot\rho$ between two circularly orbiting TianQin satellites under the influence of the Earth's static gravity field in uniform rotation. For readers' convenience, the table containing the symbols and their meanings in this paper is shown in the Appendix \ref{Appendix: symbols}.

\subsection{Basic mathematical setup}

\begin{figure}[htb]
\includegraphics[width=0.45\textwidth]{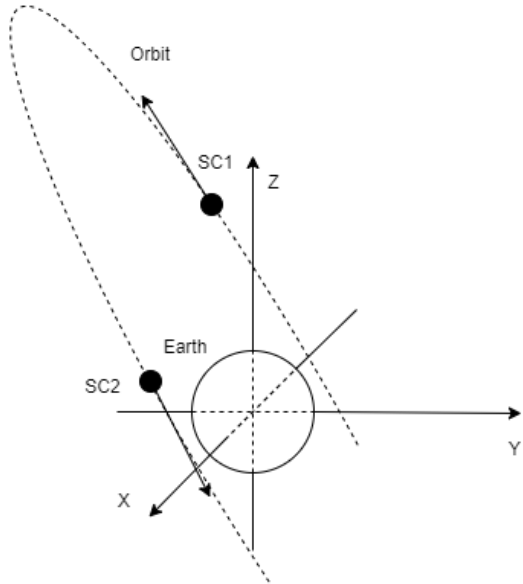}
\caption{\label{fig:satellites pair} TianQin satellites in circular prograde orbits around the Earth in the International Celestial Reference System (ICRS) with a phase difference $\gamma=120^\circ$. The orbit radius is $1\times 10^5$ km, and the inclination and longitude of the ascending node are $74.5^{\circ}$ and $211.6^{\circ}$ \cite{Ye2019}, respectively. The arrows on SC1,2 mark the flight directions. }
\end{figure}

As shown in Fig. \ref{fig:satellites pair}, we have two satellites SC1,2 moving along the same nominal circular orbit around the rotating Earth. The phase difference between the two is $\gamma=120^{\circ}$, i.e., the angle subtended by the satellites with respect to the Earth. 

In the International Terrestrial Reference System (ITRS), the Earth's gravity field can be modeled by \cite{Montenbruck2000}
\begin{equation}
\begin{split}
&V(r,\theta,\lambda)=\frac{GM}{R}\sum_{n=0}^{N}\sum_{m=0}^{n}\left(\frac{R}{r}\right)^{n+1}
\\
&\times\left(\bar{C}_{nm}\cos(m\lambda)+\bar{S}_{nm}\sin(m\lambda)\right)\bar{P}_{nm}(\cos\theta),
\end{split}
\end{equation}
where $G$ is the universal gravitational constant, $M$ the Earth's total mass, $R$ the Earth's average radius, and $n$, $m$ the degree and order of the spherical harmonic expansion, respectively. Moreover, $N$ denotes the truncation degree, and $r$, $\theta$ and $\lambda$ represent the radius, co-latitude and longitude, respectively, in the spherical coordinate system. Following the convention in gravimetry, the geo-potential takes on positive values. The spherical harmonic coefficients $\bar{C}_{nm}$ and $\bar{S}_{nm}$, and the associated Legendre functions $\bar{P}_{nm}(\cos\theta)$ are all normalized, with $\bar{C}_{1m}=0=\bar{S}_{1m}$. 

For modeling satellite motion, we switch to the International Celestial Reference System (ICRS). In order to depict the local gravity field experienced by the satellites, one may substitute the spherical coordinates with the orbital elements and express the geo-potential in terms of of the latter \cite{Kaula1961,Kaula1966}:
\begin{equation} \label{eq:ref1}
\begin{split}
&V(a,i,e,\omega,M,\Omega,\Theta)=\frac{GM}{R}\sum_{n=0}^{N}\left(\frac{R}{a}\right)^{n+1}
\\
&\times\sum_{m=0}^{n}\sum_{p=0}^{n}\bar{F}_{nmp}(i)\sum_{q=-\infty}^{\infty}G_{npq}(e)S_{nmpq}(\omega,M,\Omega,\Theta),
\end{split}
\end{equation}
where the arguments $\left\{a, i, e, \omega, M, \Omega, \Theta \right\}$ denote, respectively, the semi-major axis, the inclination, the eccentricity, the argument of perigee, the mean anomaly, the right ascension of the ascending node, and the Greenwich Apparent Sidereal Time (GAST). The explicit form of the function $S_{nmpq}$ is given by
\begin{equation} \label{eq:S_nmpq}
S_{nmpq}(\omega,M,\Omega,\Theta)=\alpha_{nm}\cos\psi_{nmpq}+\beta_{nm}\sin\psi_{nmpq}.
\end{equation}
Here the coefficients $\left\{\alpha_{nm},\beta_{nm}\right\}$ are rearrangements of $\left\{\bar{C}_{nm},\bar{S}_{nm}\right\}$: 
\begin{equation} \label{eq:alpha_beta}
\begin{array}{c}
\alpha_{nm}=\left\{
\begin{array}{cc}
\phantom{-} \bar{C}_{nm}, & n-m=\mathrm{even},\\
         -  \bar{S}_{nm}, & n-m=\mathrm{odd},
\end{array}
\right.
\\
\\
\beta_{nm}=\left\{
\begin{array}{cc}
\bar{S}_{nm}, & n-m=\mathrm{even},\\
\bar{C}_{nm}, & n-m=\mathrm{odd},
\end{array}
\right.
\end{array}
\end{equation}
and the angular argument reads
\begin{equation} \label{eq:psi_nmpq}
\psi_{nmpq}=(n-2p)\omega+(n-2p+q)M+m(\Omega-\Theta).
\end{equation}
For the function $G_{npq}(e)$, only the case of $e=0$ is used. The expression for the inclination function $\bar{F}_{nmp}(i)$ will be discussed in the next subsection, with the normalization related to $\bar{P}_{nm}(\cos\theta)$. 

It is worth mentioning that in deriving Kaula's representation Eq. (\ref{eq:ref1}), one has neglected the procession, nutation, and polar motion of the Earth's rotation axis \cite{Kaula1966}. These effects are typically slow-varying \cite{Montenbruck2000}. For TianQin's measurement band of $10^{-4}-1$ Hz (roughly 1s to 3 hours in time scales), we deem it reasonable to assume \emph{uniform} Earth rotation in the derivation and will verify it with the numerical simulation in Sec. \ref{sec:verification1}. 


\subsection{Circular reference orbits}

Now we introduce another important simplification by adopting circular reference orbits. The reasons are two-fold. First, the nominal orbit of TianQin is circular and has a large radius of $\sim 1\times 10^5$ km by design. After orbit optimization, the deviation from perfect circularity is primarily due to the Moon's and the Sun's gravitational perturbations, which increase the eccentricity only to $5\times 10^{-4}$ on average \cite{Ye2019}. Second, the approximation also makes the further derivation much more tractable, which is already quite complicated on its own. More specifically, in calculating $\ddot\rho$ (see Eq. (\ref{eq:ref3})), we will directly use the circular reference orbits and ignore the coupling between the Earth's non-spherical and third-body gravitational acceleration and the satellite's perturbed position from the circular orbit. The validity of this approximation is to be confirmed by numerical simulation in Sec. \ref{sec:verification1}. 

Assuming $e=0$, Eq. (\ref{eq:ref1}) can be simplified to \cite{Sharifi2006}
\begin{equation} \label{eq:ref2}
\begin{split}
&V=\frac{GM}{R}\sum_{n=0}^N\left(\frac{R}{a}\right)^{n+1}\sum_{m=0}^n\sum_{p=0}^n\bar{F}_{nmp}(i)
\\
&\times\left(\alpha_{nm}\cos\psi_{nmp}+\beta_{nm}\sin\psi_{nmp}\right)
\end{split}
\end{equation}
with the angular argument 
\begin{equation} \label{eq:psi_nmp}
\psi_{nmp}=(n-2p)\omega^o+m\omega_e,
\end{equation}
and
\begin{equation} \label{eq:omega-orbit-earth}
\omega^o=\omega+M, \quad \omega_e=\Omega-\Theta.
\end{equation}
Note that only the $q=0$ components of $G_{npq}(0)$ retain the value $1$ while the other vanish. Now in Eq. (\ref{eq:ref2}), the orbit elements $\{a,i,e,\Omega,\omega\}$ are of fixed values, and $M$ increases linearly with time. 

In practice, it is more convenient to rearrange the summation indices and their order so that Eq. (\ref{eq:ref2}) can be rewritten as \cite{Sharifi2006}
\begin{equation} \label{eq:ref8}
\begin{split}
&V=\sum_{m=0}^N\sum_{k=-N}^N\sum_{n=n_1[2]}^{n_2}u_n(a)\bar{F}_{nm}^k(i)
\\
&\times\left(\alpha_{nm}\cos\psi_{mk}+\beta_{nm}\sin\psi_{mk}\right),
\end{split}
\end{equation}
where one has
\begin{equation} \label{eq:function-radius}
u_n(a)=\frac{GM}{R}\left(\frac{R}{a}\right)^{n+1},
\end{equation}
and 
\begin{equation} \label{eq:psi_mk}
\psi_{mk}=m\omega_e+k\omega^o, 
\end{equation}
with $k=n-2p$. The summation bounds $\{n_1,n_2\}$ are related to $\{m,k,N\}$ by
\begin{equation} \label{eq:summation-bounds}
\begin{array}{l}
n_1=\left\{
\begin{array}{ll}
|k|, & |k|\geq m, \\
m, & |k|<m,\quad m-k=\mathrm{even}, \\
m+1, & |k|<m,\quad m-k=\mathrm{odd},
\end{array}
\right.
\\
\\
n_2=\left\{
\begin{array}{ll}
N, & N-k=\mathrm{even},  \\
N-1, & N-k=\mathrm{odd},
\end{array}
\right.
\end{array}
\end{equation}
and the notation $[2]$ means that the step size of the summation is 2. 

Introducing the auxiliary index
\begin{equation}
g=\left\{
\begin{array}{ll}
\frac{n-m}{2}, & n-m=\mathrm{even},  \\
\frac{n-m-1}{2}, & n-m=\mathrm{odd},
\end{array}
\right.
\end{equation}
one can express the inclination function $\bar{F}_{nm}^k(i)$ as
\begin{equation} \label{eq:function-inclination-another-form}
\begin{split}
&\bar{F}_{nm}^k(i)=\sum_{s=0}^{m}\sum_{t=t_1}^{t_2}\sum_{c=c_1}^{c_2}\frac{(-1)^{c+g}}{2^{2n-2t}}\frac{(2n-2t)!}{t!(n-t)!(n-m-2t)!}
\\
&\times C_m^sC_{n-m-2t+s}^cC_{m-s}^{\frac{1}{2}n-\frac{1}{2}k-t-c}(\sin i)^{n-m-2t}(\cos i)^s,
\end{split}
\end{equation}
with
\begin{equation}
t_1=0, \quad
t_2=\left\{
\begin{array}{ll}
-\frac{1}{2}k+\frac{1}{2}n, & k\ge n-2g,  \\
g, & k\le n-2g,
\end{array}
\right.
\end{equation}
and
\begin{equation}
\begin{array}{c}
c_1=\left\{
\begin{array}{cc}
0, & 2s-k-2t\le 2m-n, \\
s-\frac{1}{2}k-t+\frac{1}{2}n-m, & 2s-k-2t\ge 2m-n,
\end{array}
\right.
\\
\\
c_2=\left\{
\begin{array}{cc}
s-2t+n-m, & 2s+k-2t\le 2m-n, \\
-\frac{1}{2}k-t+\frac{1}{2}n, & 2s+k-2t\ge 2m-n.
\end{array}
\right.
\end{array}
\end{equation}
Examples of explicit forms of $\bar{F}_{nm}^k(i)$ are shown in Appendix \ref{Appendix: inclination}. 


\subsection{Range acceleration}

By differentiating the (instantaneous) range $\rho = |\mathbf{r}_2 - \mathbf{r}_1|$ between two TianQin satellites, the range acceleration in the presence of the geo-potential $V$ reads 
\begin{equation} \label{eq:ref3}
\ddot{\rho}=\left(\nabla V_2-\nabla V_1\right)\cdot\ \hat{\mathbf{e}}_{12}+\frac{1}{\rho}\left(\left|\dot{\mathbf{r}}_2-\dot{\mathbf{r}}_1\right|^2-\dot{\rho}^2\right).
\end{equation}
The first term on the right-hand side,
\begin{equation}
\mathcal{DA}:=\left(\nabla V_2-\nabla V_1\right)\cdot\hat{\mathbf{e}}_{12},
\end{equation}
represents the differential gravitational acceleration along the line of sight between the satellites, where the subscripts $\{1,2\}$ indicate SC1 and SC2, respectively, and $\hat{\mathbf{e}}_{12}$ is the unit vector from SC2 to SC1. The second term on the right-hand side of Eq. (\ref{eq:ref3}),
\begin{equation} \label{eq:ref11}
\mathcal{CA}:=\frac{1}{\rho}\left(\left|\dot{\mathbf{r}}_2-\dot{\mathbf{r}}_1\right|^2-\dot{\rho}^2\right),
\end{equation}
represents the centrifugal acceleration with $\mathbf{r}_1$, $\mathbf{r}_2$ the position vectors of SC1 and SC2, respectively. Based on the circular orbits and geometric relation, this term is given by 
\begin{equation}
\mathcal{CA}=\frac{2GM\sin(\gamma/2)}{a^2}, 
\end{equation}
which is a constant and depends on the phase difference $\gamma$. This means that in evaluating the ASD of $\ddot\rho$, we \emph{neglect} the contribution from the centrifugal acceleration, of which the validity can be verified by numerical simulation for TianQin. 

In terms of the orbit elements, the expression of $\mathcal{DA}$ can written as (see \cite{Sharifi2006} for more details)
\begin{equation} \label{eq:ref13}
\mathcal{DA}=\frac{1}{a}\left(\frac{\partial V_2}{\partial \omega^o_2}-\frac{\partial V_1}{\partial \omega^o_1}\right) \cos\left(\frac{\gamma}{2}\right) + \left(\frac{\partial V_2}{\partial r_2}+\frac{\partial V_1}{\partial r_1}\right) \sin\left(\frac{\gamma}{2}\right).
\end{equation}
Then we evaluate the partial derivatives to have
\begin{equation}
\begin{split}
&\frac{\partial V}{\partial \omega^o}=\sum_{m=0}^N\sum_{k=-N}^N\sum_{n=n_1[2]}^{n_2}ku_n(a)\bar{F}^k_{nm}(i)
\\
&\times\left(\beta_{nm}\cos\psi_{mk}-\alpha_{nm}\sin\psi_{mk}\right)
\end{split}
\end{equation}
and
\begin{equation}
\begin{split}
&\frac{\partial V}{\partial r}=\sum_{m=0}^N\sum_{k=-N}^N\sum_{n=n_1[2]}^{n_2}-\frac{n+1}{a}u_n(a)\bar{F}^k_{nm}(i)
\\
&\times\left(\alpha_{nm}\cos\psi_{mk}+\beta_{nm}\sin\psi_{mk}\right).
\end{split}
\end{equation}
By substitution into Eq. (\ref{eq:ref13}), we arrive at \cite{Sharifi2006}
\begin{equation}
\begin{split}
&\mathcal{DA}=\sum_{m=0}^N\sum_{k=-N}^N\sum_{n=n_1[2]}^{n_2}\frac{u_n(a)}{a}\bar{F}^k_{nm}(i)\times
\\
&\left[
\left(k\cos\left(\frac{\gamma}{2}\right)\beta_{nm}-(n+1)\sin\left(\frac{\gamma}{2}\right)\alpha_{nm}\right)\cos\psi_{mk}^2\right.
\\
&\!\!\!-\left(k\cos\left(\frac{\gamma}{2}\right)\alpha_{nm}+(n+1)\sin\left(\frac{\gamma}{2}\right)\beta_{nm}\right)\sin\psi_{mk}^2
\\
&\!\!\!-\left(k\cos\left(\frac{\gamma}{2}\right)\beta_{nm}+(n+1)\sin\left(\frac{\gamma}{2}\right)\alpha_{nm}\right)\cos\psi_{mk}^1
\\
&\!\!\!\left.+\left(k\cos\left(\frac{\gamma}{2}\right)\alpha_{nm}-(n+1)\sin\left(\frac{\gamma}{2}\right)\beta_{nm}\right)\sin\psi_{mk}^1
\right],
\end{split}
\end{equation}
where the superscript of $\psi_{mk}^{1,2}$ means SC1,2. Introducing $\omega^{om}$ as the argument of latitude of the midpoint of SC1,2 and
\begin{equation}
\left\{\begin{array}{l}
\psi_{mk}^1=m\omega_e+k\left(\omega^{om}-\frac{\gamma}{2}\right), \\
\psi_{mk}^2=m\omega_e+k\left(\omega^{om}+\frac{\gamma}{2}\right),
\end{array}\right.
\end{equation}
we can rewrite $\mathcal{DA}$ as \cite{Sharifi2006}
\begin{equation} \label{eq:ref9}
\begin{split}
\mathcal{DA}=\sum_{m=0}^N\sum_{k=-N}^N
&\left[E_{mk}^c\cos\left(m\omega_e+k\omega^{om}\right)\right. \\
&\!\!\!\left.+E_{mk}^s\sin\left(m\omega_e+k\omega^{om}\right)\right],
\end{split}
\end{equation}
with
\begin{eqnarray} \label{eq:E_mk-cosine-sine}
E_{mk}^c=\sum_{n=n_1[2]}^{n_2}\Xi_{nmk}(a,i,\gamma)\alpha_{nm}, \label{eq:Ec_mk} \\
E_{mk}^s=\sum_{n=n_1[2]}^{n_2}\Xi_{nmk}(a,i,\gamma)\beta_{nm}, \label{eq:Es_mk}
\end{eqnarray}
and
\begin{equation} \label{eq:Xi_nmk}
\begin{split}
&\Xi_{nmk}(a,i,\gamma)=-2\frac{u_n(a)}{a}\bar{F}_{nm}^k(i) \\
&\times\left(
k\sin\left(k\frac{\gamma}{2}\right)\cos\left(\frac{\gamma}{2}\right)
+(n+1)\sin\left(\frac{\gamma}{2}\right)\cos\left(k\frac{\gamma}{2}\right)\right).
\end{split}
\end{equation}
The variables $\omega_e$ and $\omega^{om}$ can be expressed in terms of the Earth's rotation frequency $f_e$ and the satellites' orbit frequency $f_o$, i.e.,
\begin{equation} \label{eq:ref10}
\left\{\begin{array}{l}
\omega_e=-2\pi f_et+\omega_{e0}, \\
\omega^{om}=2\pi f_ot+\omega_{o0},
\end{array}\right.
\end{equation}
where $\omega_{e0}$ and $\omega_{o0}$  are the initial phases, and $f_o$ is determined by
\begin{equation} \label{eq:frequency-satellite}
f_o=\sqrt{\frac{GM}{4 \pi^2 a^3}}.
\end{equation}
Inserting Eq. (\ref{eq:ref10}) into Eq. (\ref{eq:ref9}), we obtain
\begin{equation} \label{eq:DAE}
\mathcal{DA}=\sum_{m=0}^N\sum_{k=-N}^N
\left[
E_{mk}^c\cos\Upsilon_{mk}(t)
-E_{mk}^s\sin\Upsilon_{mk}(t)\right],
\end{equation}
with the angular argument
\begin{equation} \label{eq:Upsilon_mk}
\Upsilon_{mk}(t)=2\pi(mf_e-kf_o)t-m\omega_{e0}-k\omega_{o0}.
\end{equation}

Or equivalently, one can rewrite Eq. (\ref{eq:DAE}) as
\begin{equation} \label{eq:DAEcos}
\mathcal{DA}=\sum_{m=0}^N\sum_{k=-N}^NE_{mk}\cos\left(\Upsilon_{mk}(t)-\omega_{mk}\right),
\end{equation}
with
\begin{align} \label{eq:E_mk-omega_mk}
&E_{mk} = \sqrt{(E_{mk}^c)^2+(E_{mk}^s)^2}, \\
&\tan\omega_{mk} = -E_{mk}^s/E_{mk}^c. \label{eq:omega_mk}
\end{align}

At last, one can model the range-acceleration between two TianQin satellites by
\begin{equation} \label{eq:ran-acc}
\ddot{\rho}=\!\!\sum_{m=0}^N\sum_{k=-N}^N
\!\!\!E_{mk}\cos\left(\Upsilon_{mk}(t)-\omega_{mk}\right)+\frac{2GM\sin(\gamma/2)}{a^2},
\end{equation}
with $E_{mk}$, $\Upsilon_{mk}(t)$, and $\omega_{mk}$ given by Eqs. (\ref{eq:Ec_mk}--\ref{eq:omega_mk}). The formula consists of a series of cosine functions of time $t$. Note that terms with $m=0$ and opposite $k$ are combined into one. Their amplitudes depend on the harmonic coefficients $\bar{C}_{nm}$, $\bar{S}_{nm}$ and the satellite orbit elements $a$ and $i$, as well as the phase difference $\gamma$. The associated frequencies are integer linear combinations of the Earth's rotation frequency $f_e$ and the satellites' orbit frequency $f_o$. Hence, through a Fourier-style decomposition in terms of frequency components, the formula can reveal the frequency-domain characteristics of the effect of the Earth's gravity field \{$\bar{C}_{nm}$, $\bar{S}_{nm}$\} on the inter-satellite range-acceleration $\ddot{\rho}$. Moreover, the dependence on the orbit elements $a,i,\Omega$ also allows one to study the impact of orbit selection \cite{Luo2022}. To see explicit examples of Eq. (\ref{eq:ran-acc}), one can refer to Appendix \ref{Appendix: example A}, which contains the low-degree case of $N=2$. 


\subsection{Comments}

In Kaula's original linear perturbation method \cite{Kaula1961,Kaula1966}, the orbital element representation Eq. (\ref{eq:ref1}) of the geo-potential for a satellite moving along a reference orbit is plugged into Lagrange's perturbation equations to solve for orbital element perturbation relative to the reference values. The reference orbit is a slowly precessing circular orbit, i.e. that $\{a,e,i\}$ are of fixed values and that $\{\Omega,\omega,M\}$ are linear functions of $t$. It is obtained also by solving Lagrange's equations that contain the dominating $\bar{C}_{20}$ term. Once the orbital element perturbations are acquired, one can establish the relationship between the range or range-rate observables and the geo-potential coefficients \cite{Rosborough1986,Rosborough1987,Sharma1995}, which can be used to determine the latter from the former for gravimetry. 

Unfortunately, the solutions to Lagrange's equations containing high-degree geo-potential terms have very complicated forms, cumbersome to use for our purposes. Moreover, what we need for the noise assessment is the range-acceleration variation in the frequency domain rather than the inter-satellite range or range-rate in the time domain. Therefore, instead of solving Lagrange's equations, we directly use Kaula's representation Eq. (\ref{eq:ref1}) and calculate the range acceleration through differential gravitational acceleration sensed by the satellite pair. The treatment is inspired by \cite{Sharifi2006} and takes into account TianQin's large orbit radius and long baseline. It should be mentioned that the small orbit precession has been neglected in our analytical model, because of the large orbit radius resulting in much weaker influence of $\bar{C}_{20}$, as opposed to the case of low-orbit gravimetry missions. This has also been confirmed by the numerical work \cite{Ye2019}. 

\section{Model Verification}\label{sec:verification1}

In this section, we verify the analytical model with high-precision numerical orbit simulation. First we discuss the case of TianQin's orbit and compare the resulting ASD curves. Then we alter the orbital elements to different values \cite{Luo2022} so as to test the model in more generic settings. 

From the previous work (\cite{Zhang2021}, see Sec. IVB, Fig. 3), we have learned that the total range-acceleration ASD can be viewed as the sum of the lunisolar contribution and the Earth's contribution. Since the latter dominates above $5\times 10^{-5}$ Hz in the total ASD and hence is more relevant to TianQin, we will focus on the Earth's contribution alone in the verification. For an analytical model of the Sun and the Moon's effect on the constellation stability, one may refer to \cite{Ye2022}. 

\subsection{TianQin's orbit} \label{subsec:one group}

In this case, we set the parameters of Eq. (\ref{eq:ran-acc}) according to Table \ref{parameters} \cite{Montenbruck2000}, and use EGM2008 \cite{Pavlis2012} for the Earth's static gravity field with the maximum degree $N=12$. Then the range acceleration is calculated every 50s for a duration of 90 days. At the same time, we numerically integrate the orbits using the TQPOP program \cite{Zhang2021} with the force models listed in Table \ref{tab:models}. Note that the Earth's rotation models are also included to test our assumption of uniform Earth rotation in the model. 

\begin{table}[htb]
\caption{\label{parameters} The parameter setting of the analytical model \cite{Montenbruck2000}. }
\begin{ruledtabular}
\begin{tabular}{ccc}
Symbols & Parameters & Values \\ 
\hline
$N$ & $\text{maximum degree}$ & $12$ \\
$a$ & $\text{orbit radius}$ & $1\times10^5\mathrm{km}$ \\
$i$ & $\text{inclination}$ & $74.5^{\circ}$ \\
$\Omega$ & $\text{longitude of ascending node}$ & $211.6^{\circ}$ \\
$GM$ & $\text{Earth's gravity constant}$ & $3.986\times10^{14}\mathrm{m}^3/ \mathrm{s}^2$ \\
$R$ & $\text{average Earth radius}$ & $6.378\times10^6\mathrm{m}$ \\
$T_e$ & $\text{Earth rotation period}$ & $86164\mathrm{s}$ \\
$\omega_{o0}$ & $\text{satellite initial phase}$  & $\pi/2$ \\
\end{tabular}
\end{ruledtabular}
\end{table}

\begin{table}[htb]
\caption{\label{tab:models}
Force models implemented in the numerical simulation. }
\begin{ruledtabular}
\begin{tabular}{lc}
Models                 & Specifications \\
\hline
Earth's precession \& nutation & IAU 2006/2000A \cite{Petit2010} \\
Earth's polar motion           & EOP 14 C04 \cite{EOP} \\
Earth's static gravity field   & EGM2008 ($N=12$) \cite{Pavlis2012} \\
\end{tabular}
\end{ruledtabular}
\end{table}

\begin{figure}[htb]
\includegraphics[width=0.45\textwidth]{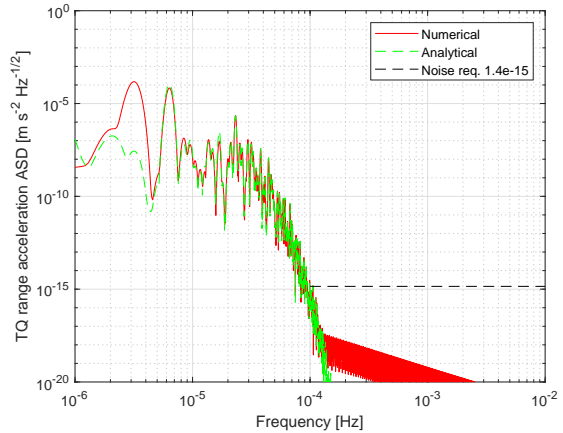}
\caption{\label{fig:ref4} Comparison of TianQin's range-acceleration ASD curves obtained from the analytical model (green) and the numerical simulation (red). The flattening of the red curve above $\sim 10^{-4}$ is due to numerical error. The mismatch at the orbital frequency $3\times 10^{-6}$ is likely owing to the assumption of the circular reference orbit in the analytical model. }
\end{figure}

The two ASD curves are compared in Fig. \ref{fig:ref4}. Note that the flattening of the red curve above $\sim 10^{-4}$ Hz is due to numerical error of TQPOP (interpolation of the EOP data, likewise for the figures in Sec. \ref{subsec:several groups}), while the green curve from Eq. (\ref{eq:ran-acc}) can extend below this error. Both curves agree well with each other except at the orbital frequency $3\times 10^{-6}$ Hz, and the much lower peak in the green curve is likely owing to the vanishing eccentricity of the circular reference orbit. However, one can see that the analytical model can indeed capture the main spectral feature of the effect due to the Earth's static gravity field. 

\subsection{Other inclinations and radii} \label{subsec:several groups}

\begin{figure}[!htp]
\includegraphics[width=0.45\textwidth]{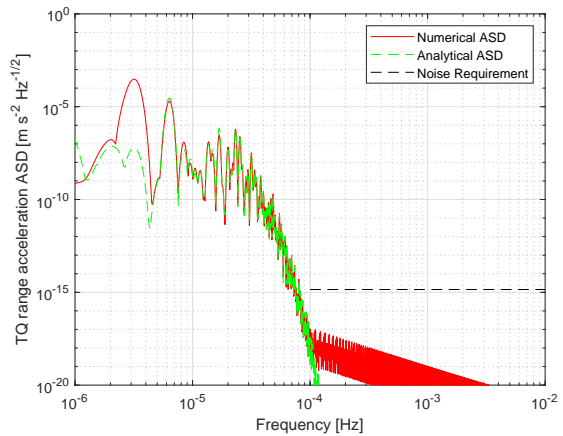}
\caption{\label{fig:ref5} Comparison of the range-acceleration ASD curves from the analytical model (green) and the numerical simulation (red) with $30^{\circ}$ inclination. }
\end{figure}

\begin{figure}[!htp]
\includegraphics[width=0.45\textwidth]{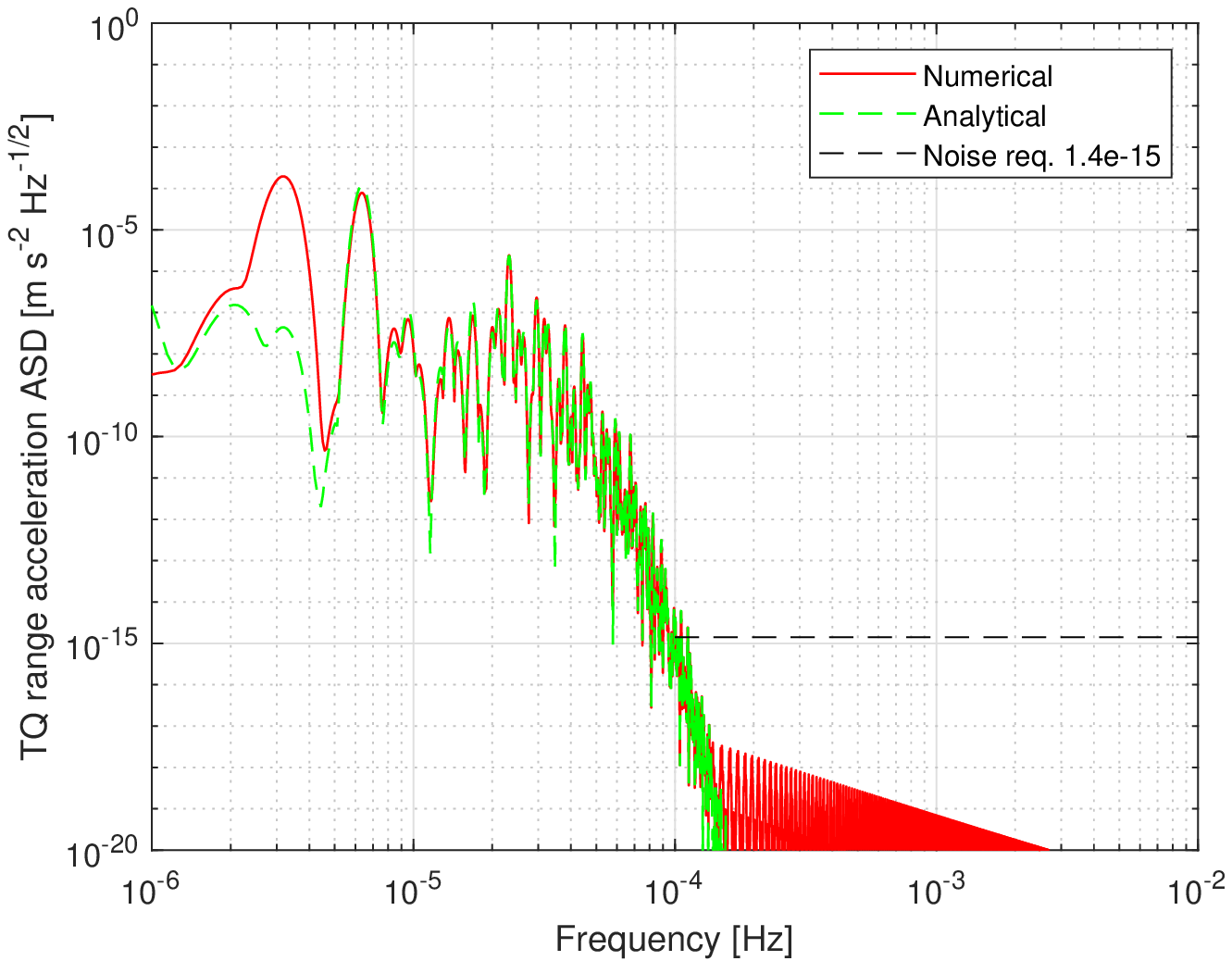}
\caption{\label{fig:ref7} Comparison of the range-acceleration ASD curves from the analytical model (green) and the numerical simulation (red) with $90^{\circ}$ inclination. }
\end{figure}

\begin{figure}[!htp]
\includegraphics[width=0.45\textwidth]{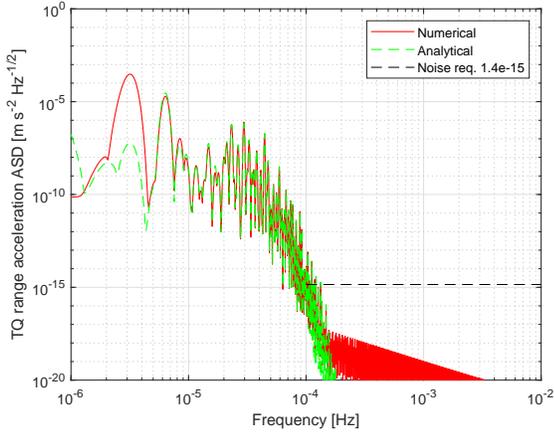}
\caption{\label{fig:ref9} Comparison of the range-acceleration ASD curves from the analytical model (green) and the numerical simulation (red) with $150^{\circ}$ inclination. }
\end{figure}

Firstly we reset the inclination to $30^{\circ}$, $90^{\circ}$, and $150^{\circ}$ in the analytical model while keeping the other parameters the same. The results are shown in Fig. \ref{fig:ref5}-\ref{fig:ref9}, all of which show good agreement and are consistent with \cite{Luo2022}. Additionally, we have also tested the cases of $60^{\circ}$ and $120^{\circ}$. They all show similar features like Fig. \ref{fig:ref5}-\ref{fig:ref9}, and hence omitted here. 

\begin{figure}[!htp]
\includegraphics[width=0.45\textwidth]{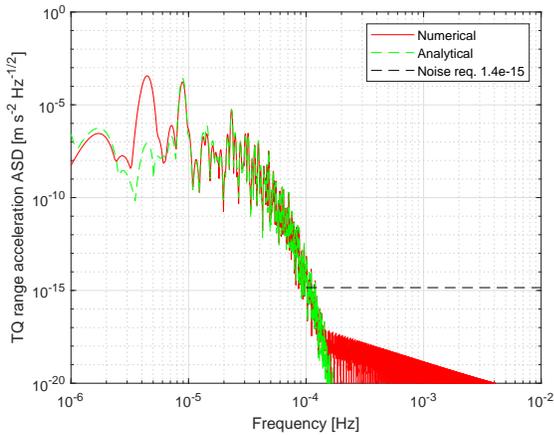}
\caption{\label{fig:ref16} Comparison of the range-acceleration ASD curves from the analytical model (green) and the numerical simulation (red) with $0.8\times10^5$km radius. }
\end{figure}

\begin{figure}[!htp]
\includegraphics[width=0.45\textwidth]{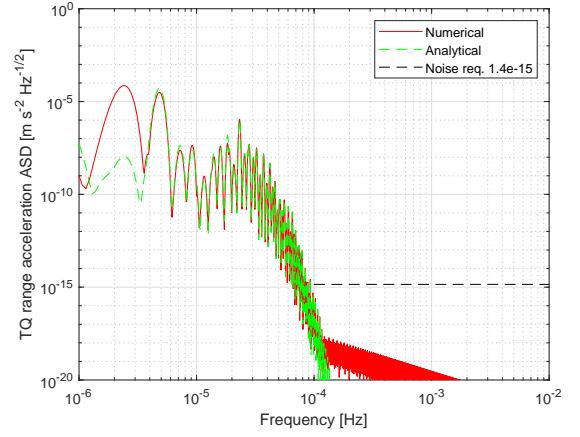}
\caption{\label{fig:ref17} Comparison of the range-acceleration ASD curves from the analytical model (green) and the numerical simulation (red) with $1.2\times10^5$km radius. }
\end{figure}

Secondly we reset the radius to $0.8\times10^5$km and $1.2\times10^5$km in the analytical model while keeping the other parameters the same. The results are shown in Fig. \ref{fig:ref16}-\ref{fig:ref17}. Again one has an evident matching of the two results. 

Due to rotational symmetry about the Earth's axis, we do not need to test different longitudes of ascending node. 
All these results have shown the validity of our model, as well as the approximations made in the derivation. It prompts us to take a further step and include time-variable gravity in the model. 


\section{Model Derivation: Time-Variable Gravity} \label{sec:expansion}

The model Eq. (\ref{eq:ran-acc}) can be extended straightforwardly to incorporate the Earth's time-variable gravity by adding time-dependent corrections to the harmonic coefficients $\bar{C}_{nm}$ and $\bar{S}_{nm}$. The temporal gravity variations can be categorized into tidal and non-tidal parts. For the Earth's tides (solid Earth, ocean, pole, and atmospheric), their analytical models in the form of the harmonic correction terms are well-known (e.g., see \cite{Petit2010}) with frequency components typically outside TianQin's measurement band of $10^{-4}-1$ Hz. As has been shown for TianQin, the tidal contributions to the total rang-acceleration ASD are considerably smaller than the static one \cite{Zhang2021}. Therefore for TianQin, it is more relevant to instead examine non-tidal gravity variations that may enter the measurement band. 

From the classification of non-tidal variations, the Earth's free oscillations are of particular interest since their frequencies are typically in the mHz range. Free oscillations of the Earth are standing waves at discrete frequencies of the solid Earth like a ringing bell (\cite{Alterman1959}, not to be confused with seismic waves). They are typically excited and observed after a large earthquake, and can last for days. In this section, we model the effect of the Earth's free oscillations triggered by earthquakes as a demonstrating example. Other types of non-tidal variations can be treated in a similar manner. 

An earthquake can be characterized by the point-source double-couple model and associated parameters such as scalar seismic moment, dip, rake, and strike \cite{Kanamori1974,Kanamori1981}. The free oscillation excited by an earthquake is a superposition of the Earth's normal modes with amplitudes determined from the parameters of the earthquake \cite{Alterman1959,Petrov2021}. The normal modes of the Earth can be calculated by the Preliminary Reference Earth Model (PREM), which is a non-rotating, elastic, spherically symmetric, one dimensional Earth model widely used in literature \cite{Dziewonski1981}. According to \cite{Ghobadi2019,Han2013}, the gravity change due to free oscillations following an earthquake can be modeled by
\begin{equation} \label{eq:dCdS}
\left\{
\begin{array}{c}
\bar{C}_{nm}(t)=\bar{C}_{nm}+\sum_{l=0}^{L} \phantom{}_{l}\Delta\bar{C}_{nm}\, \phantom{}_{l}\xi_{n}(t-t_0), \\
\bar{S}_{nm}(t)=\bar{S}_{nm}+\sum_{l=0}^{L} \phantom{}_{l}\Delta\bar{S}_{nm}\, \phantom{}_{l}\xi_{n}(t-t_0),
\end{array}
\right.
\end{equation}
where $l$ marks free-oscillation overtones, $L$ is the maximum overtone, and $t_0$ is the earthquake occurrence time. The time dependence is given by 
\begin{equation} \label{eq:xi_ln}
\phantom{}_{l}\xi_{n}(t)=
\begin{cases}
0, & t<0, \\
1-\phantom{}_l\tau_n(t)\cos\left(2\pi\phantom{}_{l}f_{n} t\right), & t\geq 0, \\
\end{cases}
\end{equation}
with the attenuation function
\begin{equation} \label{eq:attenuation-function}
\phantom{}_l\tau_n(t)=\exp\left(-\frac{2\pi\phantom{}_{l}f_{n} t}{2\phantom{}_{l}Q_{n}}\right),
\end{equation}
for each degree and overtone, where $\phantom{}_{l}f_{n}$ is the oscillation frequency, $\phantom{}_{l}Q_{n}$ is the attenuation factor, and $\phantom{}_{l}\Delta\bar{C}_{nm}$ and $\phantom{}_{l}\Delta\bar{S}_{nm}$ denote the permanent changes of the harmonic coefficients. Moreover, to be consistent with the notation of Eq. (\ref{eq:alpha_beta}), we introduce
\begin{equation} \label{eq:delta_alpha_beta}
\begin{array}{l}
\phantom{}_l\Delta\alpha_{nm}=\left\{
\begin{array}{cc}
\phantom{}_l\Delta\bar{C}_{nm}, & n-m=\mathrm{even},\\
-\phantom{}_l\Delta\bar{S}_{nm}, & n-m=\mathrm{odd},
\end{array}
\right.
\\
\\
\phantom{}_l\Delta\beta_{nm}=\left\{
\begin{array}{cc}
\phantom{}_l\Delta\bar{S}_{nm}, & n-m=\mathrm{even},\\
\phantom{}_l\Delta\bar{C}_{nm}, & n-m=\mathrm{odd},
\end{array}
\right.
\end{array}
\end{equation}
The expressions for $\alpha_{nm}(t)$ and $\beta_{nm}(t)$ can be written down accordingly. 

Including these time variations, the range-acceleration formula can be written as
\begin{equation} \label{eq:ran-acc_earthquake}
\ddot{\rho}=\mathcal{DA}_1+\mathcal{DA}_2+\mathcal{DA}_3+\mathcal{CA}.
\end{equation}
Here the first term $\mathcal{DA}_1$ is identical to $\mathcal{DA}$ in the absence of the free oscillations. The second term is given by
\begin{equation} \label{eq:permant-change}
\mathcal{DA}_2=\sum_{m=0}^N\sum_{k=-N}^NP_{mk}\cos\left(\Upsilon_{mk}(t)-\varepsilon_{mk}\right),
\end{equation}
with
\begin{equation} \label{eq:P_mk}
P_{mk}=\sqrt{\left(P_{mk}^c\right)^2+\left(P_{mk}^s\right)^2},
\end{equation}
and
\begin{equation} \label{eq:P_mk_cosine_sine}
\left\{\begin{array}{c}
P_{mk}^c=\sum_{n=n_1[2]}^{n_2}\sum_{l=0}^L\Xi_{nmk}(a,i,\gamma)\phantom{}_l\Delta\alpha_{nm}, \\
P_{mk}^s=\sum_{n=n_1[2]}^{n_2}\sum_{l=0}^L\Xi_{nmk}(a,i,\gamma)\phantom{}_l\Delta\beta_{nm},
\end{array}\right.
\end{equation}
and
\begin{equation} \label{eq:epsilon_mk}
\tan \varepsilon_{mk}=-P_{mk}^s/P_{mk}^c.
\end{equation}
The term stands for the permanent change of $\ddot{\rho}$ resulting from an earthquake, where the frequency combinations $\Upsilon_{mk}(t)$ are identical to the ones in $\mathcal{DA}$ with the amplitudes and phases slightly altered. The third term carries the free oscillation information and reads
\begin{equation} \label{eq:attentuation-oscillation-change}
\begin{split}
&\mathcal{DA}_3=\sum_{n=0}^N\sum_{l=0}^L\sum_{m=0}^n\sum_{k=-n[2]}^n\frac{1}{2}\phantom{}_lT_{nmk} \, \phantom{}_l\tau_n(t)\times \\
&\left[
\cos\left(\phantom{}_l\Psi_{nmk}^+(t)-\phantom{}_l\gamma_{nmk}\right) + \cos\left(\phantom{}_l\Psi_{nmk}^-(t)+\phantom{}_l\gamma_{nmk}\right)\right],
\end{split}
\end{equation}
with the time dependence
\begin{equation} \label{Psi_nlmk-plus-minus}
\left\{\begin{array}{l}
\begin{split}
\phantom{}_l\Psi_{nmk}^+(t)= & 2\pi\left(\phantom{}_lf_n+(mf_e-kf_o)\right)t \\
&-2\pi\phantom{}_lf_nt_0-m\omega_{e0}-k\omega_{o0},
\end{split} \\
\begin{split}
\phantom{}_l\Psi_{nmk}^-(t)= & 2\pi\left(\phantom{}_lf_n-(mf_e-kf_o)\right)t \\
&-2\pi\phantom{}_lf_nt_0+m\omega_{e0}+k\omega_{o0},
\end{split}
\end{array}\right.    
\end{equation}
and
\begin{equation} \label{eq:T_nlmk}
\phantom{}_lT_{nmk}=\sqrt{\left(\phantom{}_lT_{nmk}^c\right)^2+\left(\phantom{}_lT_{nmk}^s\right)^2},
\end{equation}
where one has
\begin{equation} \label{eq:T_nlmk-cosine-sine}
\left\{\begin{array}{l}
\phantom{}_lT_{nmk}^c=-\Xi_{nmk}(a,i,\gamma)\phantom{}_l\Delta\alpha_{nm}, \\
\phantom{}_lT_{nmk}^s=-\Xi_{nmk}(a,i,\gamma)\phantom{}_l\Delta\beta_{nm},
\end{array}\right.    
\end{equation}
and
\begin{equation} \label{eq:gamma_nlmk}
\tan \phantom{}_l\gamma_{nmk}=-\phantom{}_lT_{nmk}^s/\phantom{}_lT_{nmk}^c.
\end{equation}
The term represents a series of damped oscillations after the earthquake. Note that the resulting oscillation frequencies in $\ddot\rho$ are the free oscillation frequencies $\phantom{}_lf_n$ plus or minus the integer linear combinations of the Earth's rotation frequency $f_e$ and the satellites' orbit frequency $f_o$. This indicates \emph{splitting} of one mode frequency in the range-acceleration ASD due to the coupling with the Earth's rotation and the satellites' orbital motion. Furthermore, the amplitudes of the frequency components depend on the parameters $a,i,\gamma$ and the earthquake-induced geo-potential changes $\phantom{}_l\Delta\alpha_{nm}$ and $\phantom{}_l\Delta\beta_{nm}$. For more details, one can refer to Appendix \ref{Appendix: example B}, which shows an example of the spherical mode $\phantom{}_0S_2$ ($[n,l]=[2,0]$) oscillation in the presence of the Earth's static $N=2$ gravity field. 


\section{Model Verification} \label{sec:verification2}

Again we verify the extended model with high-precision numerical orbit simulation. Here we keep using the parameter setting from Sec. \ref{subsec:one group}. In addition, the moment magnitude of the test earthquake is 6.0 and the occurrence time $t_0$ is 30 days after the initial simulation time. The maximum degree $N$ and overtone $L$ are both set to be $12$. The mode periods $\phantom{}_lT_n=1/\phantom{}_lf_n$, the attenuation factor $\phantom{}_lQ_n$, and the geo-potential corrections $\left\{\phantom{}_l\Delta \bar{C}_{nm}, \phantom{}_l\Delta \bar{S}_{nm}\right\}$ are shown in Tables \ref{earthquake periods and attenuation factors} and \ref{earthquake coefficients} (only prominent ones are shown, e.g., $\phantom{}_0S_2, \phantom{}_0S_3, \phantom{}_1S_2, \phantom{}_1S_3$), which are obtained by following the procedures of \cite{Ghobadi2019}. 

\begin{figure}[htb]
\includegraphics[width=0.45\textwidth]{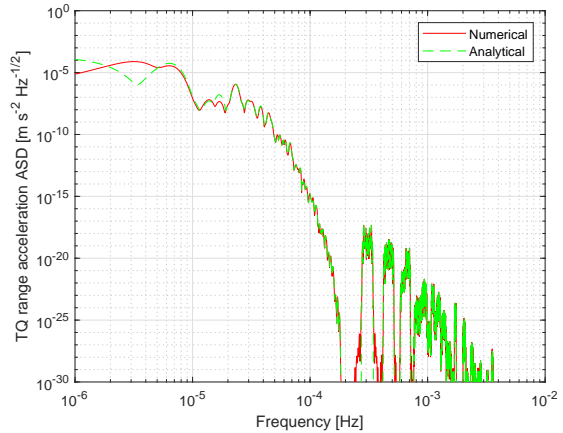}
\caption{\label{fig:ref10} 
Comparison of TianQin's range-acceleration ASD
curves obtained from the analytical model (green) and the
numerical simulation (red). The data length is 90 days and the earthquake occurs after the first 30 days. The range of the vertical axis is extended to exhibit higher-frequency free oscillation signals.}
\end{figure}

As in the previous case of static gravity, we calculate the inter-satellite range-acceleration ASD from Eq. (\ref{eq:ran-acc_earthquake}). At the same time, we obtain the numerical result from TQPOP in which the force models consist of the Earth's static gravity and the gravity change from the free oscillations mentioned above. Note that the Earth's polar motion is turned off to lower the numerical error above $\sim 10^{-4}$ due to data interpolation. The two ASD curves are compared in FIG.\ref{fig:ref10}, and the peaks above $2\times 10^{-4}$ Hz are due to the free oscillations. Both curves agree well with each other, indicating the validity of our model. The agreement also holds for other earthquakes we have tested but not shown here for succinctness. As an extra comment, it should be pointed out that spectrograms are normally preferred for visualizing non-stationary signals such as damped oscillations. Nevertheless, since our purpose is to compare the frequency content of the two results, we still use ASDs but one should keep in mind that the peak levels above $2\times 10^{-4}$ Hz do not reflect the actual amplitudes of the perturbed $\ddot{\rho}$ due to the free oscillations.

\begin{figure}[htb]
\includegraphics[width=0.45\textwidth]{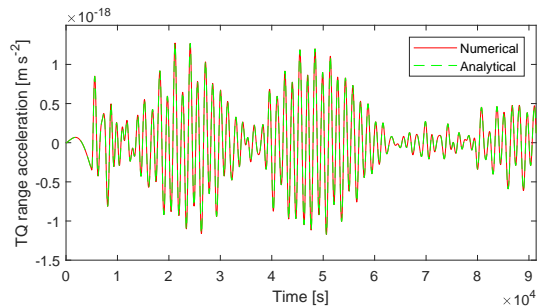}
\caption{\label{fig:ref11} 
Comparison of TianQin’s range-acceleration temporal waveforms obtained from the numerical simulation (red) and the analytical model (green). Both have undergone high-pass filtering ($>10^{-4}$ Hz) so that they reveal the influence of Earth's free oscillation induced by earthquake. The time axis above is shifted to 5000 s before the earthquake occurs. }
\end{figure}

Furthermore, we make the comparison in the time domain as well. We calculate the TianQin's range-acceleration time series using the numerical procedure and the analytical formula, respectively. Then we process the data with a high-pass filter \cite{Zheng2023} to extract signals of Earth's free oscillation induced by earthquake above $10^{-4}$ Hz. The numerical and analytical waveforms are compared in Fig.\ref{fig:ref11}. The fitting factor is $>99\%$, indicating a good consistence. 

\begin{table}[htb]
\caption{\label{earthquake periods and attenuation factors} Exemplary mode periods and attenuation factors of the tested free oscillations. }
\begin{ruledtabular}
\begin{tabular}{cccc}
Parameters & Values & Parameters & Values \\ 
\hline
$\phantom{}_0T_2$ & $3233.678\mathrm{s}$ & $\phantom{}_0Q_2$ & $509.6824$ \\
$\phantom{}_0T_3$ & $2134.577\mathrm{s}$ & $\phantom{}_0Q_3$ & $417.5499$ \\
$\cdots$ & $\cdots$ & $\cdots$ & $\cdots$ \\
$\phantom{}_1T_2$ & $1471.996\mathrm{s}$ & $\phantom{}_1Q_2$ & $310.2749$ \\
$\phantom{}_1T_3$ & $1064.875\mathrm{s}$ & $\phantom{}_1Q_3$ & $282.5018$ \\
$\cdots$ & $\cdots$ & $\cdots$ & $\cdots$ \\
\end{tabular}
\end{ruledtabular}
\end{table}

\begin{table}[htb]
\caption{\label{earthquake coefficients} Corrections to the geo-potential coefficients due to the earthquake, cf. Eq. (\ref{eq:dCdS}). }
\begin{ruledtabular}
\begin{tabular}{cccc}
Parameters & Values & Parameters & Values\\ 
\hline
$\phantom{}_0\Delta \bar{C}_{20}$ & $\phantom{+}2.954\times10^{-17}$ & $\phantom{}_0\Delta \bar{S}_{20}$ & $0$ \\
$\phantom{}_0\Delta \bar{C}_{21}$ & $\phantom{+}1.595\times10^{-16}$ & $\phantom{}_0\Delta \bar{S}_{21}$ & $-2.313\times10^{-16}$ \\
$\phantom{}_0\Delta \bar{C}_{22}$ & $\phantom{+}2.355\times10^{-16}$ & $\phantom{}_0\Delta \bar{S}_{22}$ & $\phantom{+}1.078\times10^{-16}$ \\
$\phantom{}_0\Delta \bar{C}_{30}$ & $-2.118\times10^{-17}$ & $\phantom{}_0\Delta \bar{S}_{30}$ & $0$ \\
$\phantom{}_0\Delta \bar{C}_{31}$ & $-9.627\times10^{-16}$ & $\phantom{}_0\Delta \bar{S}_{31}$ & $\phantom{+}6.953\times10^{-16}$ \\
$\phantom{}_0\Delta \bar{C}_{32}$ & $\phantom{+}6.099\times10^{-17}$ & $\phantom{}_0\Delta \bar{S}_{32}$ & $\phantom{+}6.132\times10^{-17}$ \\
$\phantom{}_0\Delta \bar{C}_{33}$ & $\phantom{+}3.797\times10^{-16}$ & $\phantom{}_0\Delta \bar{S}_{33}$ & $\phantom{+}8.375\times10^{-16}$ \\
$\cdots$ & $\cdots$ & $\cdots$ & $\cdots$ \\
$\phantom{}_1\Delta \bar{C}_{20}$ & $-1.896\times10^{-17}$ & $\phantom{}_1\Delta \bar{S}_{20}$ & $0$ \\
$\phantom{}_1\Delta \bar{C}_{21}$ & $\phantom{+}1.104\times10^{-15}$ & $\phantom{}_1\Delta \bar{S}_{21}$ & $-8.167\times10^{-16}$ \\
$\phantom{}_1\Delta \bar{C}_{22}$ & $\phantom{+}1.269\times10^{-15}$ & $\phantom{}_1\Delta \bar{S}_{22}$ & $\phantom{+}3.188\times10^{-16}$ \\
$\phantom{}_1\Delta \bar{C}_{30}$ & $\phantom{+}2.772\times10^{-18}$ & $\phantom{}_1\Delta \bar{S}_{30}$ & $0$ \\
$\phantom{}_1\Delta \bar{C}_{31}$ & $\phantom{+}1.392\times10^{-15}$ & $\phantom{}_1\Delta \bar{S}_{31}$ & $-1.043\times10^{-15}$ \\
$\phantom{}_1\Delta \bar{C}_{32}$ & $-9.928\times10^{-17}$ & $\phantom{}_1\Delta \bar{S}_{32}$ & $-4.501\times10^{-17}$ \\
$\phantom{}_1\Delta \bar{C}_{33}$ & $-5.365\times10^{-16}$ & $\phantom{}_1\Delta \bar{S}_{33}$ & $-1.237\times10^{-15}$ \\
$\cdots$ & $\cdots$ & $\cdots$ & $\cdots$ \\
\end{tabular}
\end{ruledtabular}
\end{table}

\section{Concluding Remarks and Outlook}\label{sec:conclusions}

In this work, we provide an analytical model which can account for the perturbing effect of the Earth's gravity field on TianQin's inter-satellite range acceleration. Especially, we have verified that the ASD curves obtained from the analytical model matches those from the numerical simulation. The work helps to confirm the numerical result of \cite{Zhang2021} and to better understand the underlying mechanism in terms of the frequency composition, as well as the dependence on orbit selection \cite{Luo2022}. Particularly, we have also extended the model to capture the time-variable gravity from the Earth's free oscillations, which enters TianQin's measurement band. 

For future applications in environment monitoring and noise-reduction pipelines, the model is useful in studying TianQin's detector response to gravity disturbance from the Earth, particularly through time delay interferometry combinations \cite{Tinto2021}. Additionally, the model may also find usage in fast waveform generation for subtracting unwanted terrestrial gravity disturbance in case of occasional large-scale geo-seismic activities. Some related work will be reported in the future. 


\begin{acknowledgements}
The authors thank Hang Li, Xiang'e Lei, and Jun Luo for helpful discussions and comments, and special thanks to Kun Liu for providing earthquake parameters. X. Z. is supported by the National Key R\&D Program of China (Grant No. 2020YFC2201202 and and 2022YFC2204600). 
\end{acknowledgements}


\appendix

\section{Table of Symbols}\label{Appendix: symbols}

Table \ref{tab:symbols} below lists the symbols used in the paper as well as their meanings for quick look-ups.

\begin{longtable*}{@{\extracolsep{0in}}p{3in}p{4in}}
\caption{\label{tab:symbols} List of symbols and their meanings}\\*
\noalign{\vspace{3pt}}%
\toprule\rule{0pt}{12pt}
\textbf{Symbols}&\textbf{Meanings}\\*[3pt]
\endfirsthead
\multicolumn{2}{c}{TABLE~\ref{tab:symbols} (continued): Symbols and Meanings}%
\rule{0pt}{12pt}\\[3pt]
\colrule\rule{0pt}{12pt}
\textbf{Symbols}&\textbf{Meanings}\\*[3pt]
\endhead
\noalign{\nobreak\vspace{3pt}}%
\colrule
\endfoot
\noalign{\nobreak\vspace{3pt}}%
\botrule
\endlastfoot
$V$ & Earth's gravity potential \\
$G$ & universal gravitational constant \\
$M$ & Earth's total mass \\
$R$ & Earth's average radius \\
$n$ & degree of spherical harmonic expansion \\
$m$ & order of spherical harmonic expansion \\
$N$ & truncation degree \\
$l$ & Earth's free-oscillation overtone \\
$L$ & maximum overtone \\
$p$ & index with values $0,1,\cdots,n$ \\
$q$ & index with values $-\infty,\cdots,-1,0,1,\cdots,+\infty$ \\
$k$ & index with values $-N,\cdots,-1,0,1,\cdots,N$ \\
$n_1,n_2$ & see Eq.(\ref{eq:summation-bounds}) \\
$t$ & time \\
$t_0$ &  earthquake occurrence time \\
$r$ & radius \\
$\theta$ & co-latitude \\
$\lambda$ & longitude \\
$\gamma$ & phase difference between two satellites \\
$a$ & semi-major axis \\
$i$ & inclination \\
$e$ & eccentricity \\
$\omega$ & argument of perigee \\
$M$ & mean anomaly \\
$\Omega$ & right ascension of ascending node \\
$\Theta$ & Greenwich Apparent Sidereal Time \\
$\bar{C}_{nm},\bar{S}_{nm}$ & spherical harmonic coefficients \\
$\alpha_{nm}$,$\beta_{nm}$ & rearrangements of $\bar{C}_{nm}$,$\bar{S}_{nm}$ defined by Eq.(\ref{eq:alpha_beta}) \\
$\phantom{}_{l}\Delta\bar{C}_{nm},\phantom{}_{l}\Delta\bar{S}_{nm}$ & coefficients of correction of the harmonic coefficients \\
$\phantom{}_l\Delta\alpha_{nm},\phantom{}_l\Delta\beta_{nm}$ & rearrangement of $\phantom{}_{l}\Delta\bar{C}_{nm},\phantom{}_{l}\Delta\bar{S}_{nm}$ defined by Eq.(\ref{eq:delta_alpha_beta}) \\
$\bar{P}_{nm}(\cos\theta)$ & normalized associated Legendre functions \\
$S_{nmpq}(\omega,M,\Omega,\Theta)$ & see Eq.(\ref{eq:S_nmpq}) \\
$\psi_{nmpq}$ & see Eq.(\ref{eq:psi_nmpq}) \\
$\psi_{nmp}$ & see Eq.(\ref{eq:psi_nmp}) \\
$\psi_{mk}$ & see Eq.(\ref{eq:psi_mk}) \\
$G_{npq}(e)$ & functions of eccentricity \\
$\bar{F}_{nmp}(i)$ & functions of inclination \\
$\bar{F}_{nm}^k(i)$ & another form of functions of inclination defined by Eq.(\ref{eq:function-inclination-another-form}) \\
$u_n(a)$ & functions of radius defined by Eq.(\ref{eq:function-radius}) \\
$\omega^o$ & see Eq.(\ref{eq:omega-orbit-earth}) \\
$\omega_e$ & see Eq.(\ref{eq:omega-orbit-earth}) \\
$\omega^{om}$ & argument of latitude of the midpoint of SC1,2 \\
$f_e$ & Earth's rotation frequency \\
$f_o$ & satellites' orbit frequency obtained from Eq.(\ref{eq:frequency-satellite}) \\
$\omega_{e0}$ & initial phases of Earth's rotation \\
$\omega_{o0}$ & initial phases of satellites' movement \\
$\phantom{}_{l}f_{n}$ & Earth's free-oscillation frequency \\
$\phantom{}_{l}Q_{n}$ & Earth's free-oscillation attenuation factor \\
$E_{mk}$ & see Eq.(\ref{eq:E_mk-omega_mk}) \\
$E_{mk}^c,E_{mk}^s$ & see Eq.(\ref{eq:E_mk-cosine-sine}) \\
$\omega_{mk}$ & see Eq.(\ref{eq:E_mk-omega_mk}) \\
$P_{mk}$ & see Eq.(\ref{eq:P_mk}) \\
$P_{mk}^c,P_{mk}^s$ & see Eq.(\ref{eq:P_mk_cosine_sine}) \\
$\varepsilon_{mk}$ & see Eq.(\ref{eq:epsilon_mk}) \\
$\phantom{}_lT_{nmk}$ & see Eq.(\ref{eq:T_nlmk}) \\
$\phantom{}_lT_{nmk}^c,\phantom{}_lT_{nmk}^s$ & see Eq.(\ref{eq:T_nlmk-cosine-sine}) \\
$\phantom{}_l\gamma_{nmk}$ & see Eq.(\ref{eq:gamma_nlmk}) \\
$\Xi_{nmk}(a,i,\gamma)$ & see Eq.(\ref{eq:Xi_nmk}) \\
$\Upsilon_{mk}(t)$ & see Eq.(\ref{eq:Upsilon_mk}) \\
$\phantom{}_{l}\xi_{n}(t)$ & see Eq.(\ref{eq:xi_ln}) \\
$\phantom{}_l\tau_n(t)$ & attenuation function defined by Eq.(\ref{eq:attenuation-function}) \\
$\phantom{}_l\Psi_{nmk}^+(t),\phantom{}_l\Psi_{nmk}^-(t)$ & see Eq.(\ref{Psi_nlmk-plus-minus}) \\
$\mathcal{CA}$ & centrifugal acceleration \\
$\mathcal{DA}$ & projected differential gravitational acceleration \\
$\mathcal{DA}_1$ & identical to $\mathcal{DA}$ \\
$\mathcal{DA}_2$ & permanent change of $\ddot{\rho}$ given by Eq.(\ref{eq:permant-change}) \\
$\mathcal{DA}_3$ & attenuation oscillation change of $\ddot{\rho}$ given by Eq.(\ref{eq:attentuation-oscillation-change}) \\
$\rho$ & inter-satellite range \\
$\dot\rho$ & inter-satellite range-rate \\
$\ddot\rho$ & inter-satellite range-acceleration \\
$\mathbf{r}_1,\mathbf{r}_2$ & position vectors of SC1 and SC2 \\
$\dot{\mathbf{r}}_1,\dot{\mathbf{r}}_2$ & velocity vectors of SC1 and SC2 \\
$\hat{\mathbf{e}}_{12}$ & unit vector from SC2 to SC1 \\
\end{longtable*}


\section{Explicit forms of inclination functions}\label{Appendix: inclination}
A non-normalized version of $\bar{F}_{nmp}$ up to $n=4$ can be found in \cite{Kaula1966}. The explicit forms of normalized $\bar{F}_{nm}^k(i)$ up to $n=4$ used in this work are shown below, following the ascending order of $[m,k,n]$. 

\begin{equation}
\bar{F}_{4\,0}^{-4}(i)=\frac{105}{128}(\sin i)^4
\end{equation}
\begin{equation}
\bar{F}_{3\,0}^{-3}(i)=\frac{5}{16}\sqrt{7}(\sin i)^3 
\end{equation}
\begin{equation}
\bar{F}_{2\,0}^{-2}(i)=-\frac{3}{8}\sqrt{5}(\sin i)^2
\end{equation}
\begin{equation}
\bar{F}_{4\,0}^{-2}(i)=-\frac{105}{32}(\sin i)^4+\frac{45}{16}(\sin i)^2
\end{equation}
\begin{equation}
\bar{F}_{1\,0}^{-1}(i)=-\frac{1}{2}\sqrt{3}\sin i
\end{equation}
\begin{equation}
\bar{F}_{3\,0}^{-1}(i)=-\frac{15}{16}\sqrt{7}(\sin i)^3+\frac{3}{4}\sqrt{7}\sin i
\end{equation}
\begin{equation}
\bar{F}_{0\,0}^{0}(i)=1
\end{equation}
\begin{equation}
\bar{F}_{2\,0}^{0}(i)=\frac{3}{4}\sqrt{5}(\sin i)^2-\frac{1}{2}\sqrt{5}
\end{equation}
\begin{equation}
\bar{F}_{4\,0}^{0}(i)=\frac{315}{64}(\sin i)^4-\frac{45}{8}(\sin i)^2+\frac{9}{8}
\end{equation}
\begin{equation}
\bar{F}_{1\,0}^{1}(i)=\frac{1}{2}\sqrt{3}\sin i
\end{equation}
\begin{equation}
\bar{F}_{3\,0}^{1}(i)=\frac{15}{16}\sqrt{7}(\sin i)^3-\frac{3}{4}\sqrt{7}\sin i
\end{equation}
\begin{equation}
\bar{F}_{2\,0}^{2}(i)=-\frac{3}{8}\sqrt{5}(\sin i)^2
\end{equation}
\begin{equation}
\bar{F}_{4\,0}^{2}(i)=-\frac{105}{32}(\sin i)^4+\frac{45}{16}(\sin i)^2
\end{equation}
\begin{equation}
\bar{F}_{3\,0}^{3}(i)=-\frac{5}{16}\sqrt{7}(\sin i)^3 
\end{equation}
\begin{equation}
\bar{F}_{4\,0}^{4}(i)=\frac{105}{128}(\sin i)^4
\end{equation}
\begin{equation}
\bar{F}_{4\,1}^{-4}(i)=-\frac{21}{64}\sqrt{10}(\sin i)^3\cos i+\frac{21}{64}\sqrt{10}(\sin i)^3
\end{equation}
\begin{equation}
\bar{F}_{3\,1}^{-3}(i)=\frac{5}{32}\sqrt{42}(\sin i)^2\cos i-\frac{5}{32}\sqrt{42}(\sin i)^2
\end{equation}
\begin{equation}
\bar{F}_{2\,1}^{-2}(i)=\frac{1}{4}\sqrt{15}\sin i\cos i-\frac{1}{4}\sqrt{15}\sin i
\end{equation}
\begin{equation}
\begin{split}
\bar{F}_{4\,1}^{-2}(i)&=\frac{21}{16}\sqrt{10}(\sin i)^3\cos i-\frac{9}{16}\sqrt{10}\sin i\cos i
\\
&-\frac{21}{32}\sqrt{10}(\sin i)^3+\frac{9}{16}\sqrt{10}\sin i
\end{split}
\end{equation}
\begin{equation}
\bar{F}_{1\,1}^{-1}(i)=-\frac{1}{2}\sqrt{3}\cos i+\frac{1}{2}\sqrt{3}
\end{equation}
\begin{equation}
\begin{split}
\bar{F}_{3\,1}^{-1}(i)=&-\frac{15}{32}\sqrt{42}(\sin i)^2\cos i+\frac{5}{32}\sqrt{42}(\sin i)^2
\\
&+\frac{1}{8}\sqrt{42}\cos i-\frac{1}{8}\sqrt{42}
\end{split}
\end{equation}
\begin{equation}
\bar{F}_{2\,1}^{0}(i)=-\frac{1}{2}\sqrt{15}\sin i\cos i
\end{equation}
\begin{equation}
\bar{F}_{4\,1}^{0}(i)=-\frac{63}{32}\sqrt{10}(\sin i)^3\cos i+\frac{9}{8}\sqrt{10}\sin i\cos i
\end{equation}
\begin{equation}
\bar{F}_{1\,1}^{1}(i)=\frac{1}{2}\sqrt{3}\cos i+\frac{1}{2}\sqrt{3}
\end{equation}
\begin{equation}
\begin{split}
\bar{F}_{3\,1}^{1}(i)&=\frac{15}{32}\sqrt{42}(\sin i)^2\cos i+\frac{5}{32}\sqrt{42}(\sin i)^2
\\
&-\frac{1}{8}\sqrt{42}\cos i-\frac{1}{8}\sqrt{42}
\end{split}
\end{equation}
\begin{equation}
\bar{F}_{2\,1}^{2}(i)=\frac{1}{4}\sqrt{15}\sin i\cos i+\frac{1}{4}\sqrt{15}\sin i
\end{equation}
\begin{equation}
\begin{split}
\bar{F}_{4\,1}^{2}(i)&=\frac{21}{16}\sqrt{10}(\sin i)^3\cos i-\frac{9}{16}\sqrt{10}\sin i\cos i
\\
&+\frac{21}{32}\sqrt{10}(\sin i)^3-\frac{9}{16}\sqrt{10}\sin i
\end{split}
\end{equation}
\begin{equation}
\bar{F}_{3\,1}^{3}(i)=-\frac{5}{32}\sqrt{42}(\sin i)^2\cos i-\frac{5}{32}\sqrt{42}(\sin i)^2
\end{equation}
\begin{equation}
\bar{F}_{4\,1}^{4}(i)=-\frac{21}{64}\sqrt{10}(\sin i)^3\cos i-\frac{21}{64}\sqrt{10}(\sin i)^3
\end{equation}
\begin{equation}
\begin{split}
\bar{F}_{4\,2}^{-4}(i)=&-\frac{21}{64}\sqrt{5}(\sin i)^2(\cos i)^2+\frac{21}{32}\sqrt{5}(\sin i)^2\cos i
\\
&-\frac{21}{64}\sqrt{5}(\sin i)^2
\end{split}
\end{equation}
\begin{equation}
\begin{split}
\bar{F}_{3\,2}^{-3}(i)=&-\frac{1}{16}\sqrt{105}\sin i(\cos i)^2+\frac{1}{8}\sqrt{105}\sin i\cos i
\\
&-\frac{1}{16}\sqrt{105}\sin i
\end{split}
\end{equation}
\begin{equation}
\bar{F}_{2\,2}^{-2}(i)=\frac{1}{8}\sqrt{15}(\cos i)^2-\frac{1}{4}\sqrt{15}\cos i+\frac{1}{8}\sqrt{15}
\end{equation}
\begin{equation}
\begin{split}
\bar{F}_{4\,2}^{-2}(i)&=\frac{21}{16}\sqrt{5}(\sin i)^2(\cos i)^2-\frac{21}{16}\sqrt{5}(\sin i)^2\cos i
\\
&-\frac{3}{16}\sqrt{5}(\cos i)^2+\frac{3}{8}\sqrt{5}\cos i-\frac{3}{16}\sqrt{5}
\end{split}
\end{equation}
\begin{equation}
\begin{split}
\bar{F}_{3\,2}^{-1}(i)&=\frac{3}{16}\sqrt{105}\sin i(\cos i)^2-\frac{1}{8}\sqrt{105}\sin i\cos i
\\
&-\frac{1}{16}\sqrt{105}\sin i
\end{split}
\end{equation}
\begin{equation}
\bar{F}_{2\,2}^{0}(i)=-\frac{1}{4}\sqrt{15}(\cos i)^2+\frac{1}{4}\sqrt{15}
\end{equation}
\begin{equation}
\begin{split}
\bar{F}_{4\,2}^{0}(i)=&-\frac{63}{32}\sqrt{5}(\sin i)^2(\cos i)^2+\frac{21}{32}\sqrt{5}(\sin i)^2
\\
&+\frac{3}{8}\sqrt{5}(\cos i)^2-\frac{3}{8}\sqrt{5}
\end{split}
\end{equation}
\begin{equation}
\begin{split}
\bar{F}_{3\,2}^{1}(i)=&-\frac{3}{16}\sqrt{105}\sin i(\cos i)^2-\frac{1}{8}\sqrt{105}\sin i\cos i
\\
&+\frac{1}{16}\sqrt{105}\sin i
\end{split}
\end{equation}
\begin{equation}
\bar{F}_{2\,2}^{2}(i)=\frac{1}{8}\sqrt{15}(\cos i)^2+\frac{1}{4}\sqrt{15}\cos i+\frac{1}{8}\sqrt{15}
\end{equation}
\begin{equation}
\begin{split}
\bar{F}_{4\,2}^{2}(i)&=\frac{21}{16}\sqrt{5}(\sin i)^2(\cos i)^2+\frac{21}{16}\sqrt{5}(\sin i)^2\cos i
\\
&-\frac{3}{16}\sqrt{5}(\cos i)^2-\frac{3}{8}\sqrt{5}\cos i-\frac{3}{16}\sqrt{5}
\end{split}
\end{equation}
\begin{equation}
\begin{split}
\bar{F}_{3\,2}^{3}(i)&=\frac{1}{16}\sqrt{105}\sin i(\cos i)^2+\frac{1}{8}\sqrt{105}\sin i\cos i
\\
&+\frac{1}{16}\sqrt{105}\sin i
\end{split}
\end{equation}
\begin{equation}
\begin{split}
\bar{F}_{4\,2}^{4}(i)=&-\frac{21}{64}\sqrt{5}(\sin i)^2(\cos i)^2-\frac{21}{32}\sqrt{5}(\sin i)^2\cos i
\\
&-\frac{21}{64}\sqrt{5}(\sin i)^2
\end{split}
\end{equation}


\section{An explicit example for the Earth's static gravity field} \label{Appendix: example A}

To help gain more intuition about Eq. (\ref{eq:ran-acc}), we exhibit its explicit form for $N=2$ static gravity field. In this case, one has
\begin{equation} \label{eq:example}
\begin{split}
\ddot{\rho}&=2GM\sin(\gamma/2)/a^2+A_{00}+A_{02}\cos\left(2\pi f_{02}t+\varphi_{02}\right) \\
&+A_{10}\cos\left(2\pi f_{10}t+\varphi_{10}\right)+A_{20}\cos\left(2\pi f_{20}t+\varphi_{20}\right) \\
&+A_{1-2}\cos\left(2\pi f_{1-2}t+\varphi_{1-2}\right)+A_{12}\cos\left(2\pi f_{12}t+\varphi_{12}\right) \\
&+A_{2-2}\cos\left(2\pi f_{2-2}t+\varphi_{2-2}\right)+A_{22}\cos\left(2\pi f_{22}t+\varphi_{22}\right).
\end{split}
\end{equation}
As can be seen, besides the constant terms, there are several cosine terms of the time $t$ in Eq. (\ref{eq:example}). The amplitudes, frequencies and phases of these cosine term are given below. 
\newline
(1) Term with the frequency 0:
\begin{equation}
\begin{split}
A_{00}=&-2\frac{GM}{a^2}\sin \left(\frac{\gamma}{2}\right)
\\
&-\frac{3\sqrt{5}}{2}\frac{GM}{R^2}\left(\frac{R}{a}\right)^4\left(3(\sin i)^2-2\right)\sin \left(\frac{\gamma}{2}\right)\bar{C}_{20}.
\end{split}
\end{equation}
(2) Term with the frequency $f_{02}$:
\begin{equation}
f_{02}=2f_o,
\end{equation}
\begin{equation}
\begin{split}
&A_{02}=\frac{3\sqrt{5}}{2}\frac{GM}{R^2}\left(\frac{R}{a}\right)^4(\sin i)^2
\\
&\times\left(2\sin \gamma \cos \left(\frac{\gamma}{2}\right)+3\sin \left(\frac{\gamma}{2}\right)\cos \gamma\right)\bar{C}_{20},
\end{split}
\end{equation}
\begin{equation}
\varphi_{02}=2\omega_{o0}.
\end{equation}
(3) Term with the frequency $f_{10}$:
\begin{equation}
f_{10}=f_e,
\end{equation}
\begin{equation}
A_{10}=\sqrt{\left(E_{10}^c\right)^2+\left(E_{10}^s\right)^2},
\end{equation}
\begin{equation}
\left\{\begin{array}{c}
E_{10}^c=-3\sqrt{15}\frac{GM}{R^2}\left(\frac{R}{a}\right)^4\sin i\cos i\sin \left(\frac{\gamma}{2}\right)\bar{S}_{21}, \\
E_{10}^s=+3\sqrt{15}\frac{GM}{R^2}\left(\frac{R}{a}\right)^4\sin i\cos i\sin \left(\frac{\gamma}{2}\right)\bar{C}_{21},
\end{array}\right.
\end{equation}
\begin{equation}
\varphi_{10}=-\omega_{e0}-\omega_{10},
\end{equation}
\begin{equation}
\tan\omega_{10}=-E_{10}^s/E_{10}^c.
\end{equation}
(4) Term with the frequency $f_{20}$:
\begin{equation}
f_{20}=2f_e,
\end{equation}
\begin{equation}
A_{20}=\sqrt{\left(E_{20}^c\right)^2+\left(E_{20}^s\right)^2},
\end{equation}
\begin{equation}
\left\{\begin{array}{c}
E_{20}^c=\frac{3\sqrt{15}}{2}\frac{GM}{R^2}\left(\frac{R}{a}\right)^4\left((\cos i)^2-1\right)\sin \left(\frac{\gamma}{2}\right)\bar{C}_{22}, \\
E_{20}^s=\frac{3\sqrt{15}}{2}\frac{GM}{R^2}\left(\frac{R}{a}\right)^4\left((\cos i)^2-1\right)\sin \left(\frac{\gamma}{2}\right)\bar{S}_{22},
\end{array}\right.
\end{equation}
\begin{equation}
\varphi_{20}=-2\omega_{e0}-\omega_{20},
\end{equation}
\begin{equation}
\tan\omega_{20}=-E_{20}^s/E_{20}^c.
\end{equation}
(5) Term with the frequency $f_{1\mp2}$:
\begin{equation}
f_{1\mp2}=f_e\pm2f_o,
\end{equation}
\begin{equation}
A_{1\mp2}=\sqrt{\left(E_{1\mp2}^c\right)^2+\left(E_{1\mp2}^s\right)^2},
\end{equation}
\begin{equation}
\left\{\begin{array}{c}
\begin{split}
&E_{1\mp2}^c=+\frac{\sqrt{15}}{2}\frac{GM}{R^2}\left(\frac{R}{a}\right)^4\sin i\left(\cos i\mp1\right) \\
&\times\left(2\sin \gamma \cos \left(\frac{\gamma}{2}\right)+3\sin \left(\frac{\gamma}{2}\right)\cos \gamma\right)\bar{S}_{21},
\end{split}
\\
\begin{split}
&E_{1\mp2}^s=-\frac{\sqrt{15}}{2}\frac{GM}{R^2}\left(\frac{R}{a}\right)^4\sin i\left(\cos i\mp1\right) \\
&\times\left(2\sin \gamma \cos \left(\frac{\gamma}{2}\right)+3\sin \left(\frac{\gamma}{2}\right)\cos \gamma\right)\bar{C}_{21},
\end{split}
\end{array}\right.
\end{equation}
\begin{equation}
\varphi_{1\mp2}=-\omega_{e0}\pm2\omega_{o0}-\omega_{1\mp2},
\end{equation}
\begin{equation}
\tan\omega_{1\mp2}=-E_{1\mp2}^s/E_{1\mp2}^c.
\end{equation}
(6) Term with the frequency $f_{2\mp2}$:
\begin{equation}
f_{2\mp2}=2f_e\pm2f_o,
\end{equation}
\begin{equation}
A_{2\mp2}=\sqrt{\left(E_{2\mp2}^c\right)^2+\left(E_{2\mp2}^s\right)^2},
\end{equation}
\begin{equation}
\left\{\begin{array}{c}
\begin{split}
&E_{2\mp2}^c=-\frac{\sqrt{15}}{4}\frac{GM}{R^2}\left(\frac{R}{a}\right)^4\left(\cos i\mp1\right)^2 \\ 
&\times\left(2\sin \gamma \cos \left(\frac{\gamma}{2}\right)+3\sin \left(\frac{\gamma}{2}\right)\cos \gamma\right)\bar{C}_{22},
\end{split}
\\
\begin{split}
&E_{2\mp2}^s=-\frac{\sqrt{15}}{4}\frac{GM}{R^2}\left(\frac{R}{a}\right)^4\left(\cos i\mp1\right)^2 \\ 
&\times\left(2\sin \gamma \cos \left(\frac{\gamma}{2}\right)+3\sin \left(\frac{\gamma}{2}\right)\cos \gamma\right)\bar{S}_{22},
\end{split}
\end{array}\right.
\end{equation}
\begin{equation}
\varphi_{1\mp2}=-2\omega_{e0}\pm2\omega_{o0}-\omega_{2\mp2},
\end{equation}
\begin{equation}
\tan\omega_{2\mp2}=-E_{2\mp2}^s/E_{2\mp2}^c.
\end{equation}
The resulting frequencies are integer linear combinations of the Earth's rotation frequency $f_e$ and the satellites' orbital frequency $f_o$, i.e., 2$f_o$, $f_e$, 2$f_e$, $f_e\mp2f_o$, $2f_e\mp2f_o$. The amplitudes are functions of the Earth's gravity field coefficients $\bar{C}_{20}$, $\bar{C}_{21}$, $\bar{S}_{21}$, $\bar{C}_{22}$, $\bar{S}_{22}$ and TianQin's orbital parameters $a$, $i$, $\gamma$. The phases are determined by the initial phases of the Earth's rotation and the satellite orbits, $\omega_{e0}$ and $\omega_{o0}$, as well as the parameters $a$, $i$, $\gamma$. The frequencies and associated amplitudes correspond to the peaks of the ASD curves. 


\section{An explicit example for the Earth's time-variable gravity field} \label{Appendix: example B}

To help gain more intuition about about Eq. (\ref{eq:ran-acc_earthquake}), we exhibit its explicit form for $N=2$ static gravity field with the oscillation mode $\phantom{}_0S_2$ ($[n,l]=[2,0]$). After the earthquake's occurrence, the $N=2$ terms change into
\begin{equation}
\left\{
\begin{array}{c}
\bar{C}_{20}(t)=\bar{C}_{20}+\phantom{}_{0}\Delta\bar{C}_{20}\, \phantom{}_{0}\xi_{2}(t-t_0),
\\
\bar{C}_{21}(t)=\bar{C}_{21}+\phantom{}_{0}\Delta\bar{C}_{21}\, \phantom{}_{0}\xi_{2}(t-t_0),
\\
\bar{S}_{21}(t)=\bar{S}_{21}+\phantom{}_{0}\Delta\bar{S}_{21}\, \phantom{}_{0}\xi_{2}(t-t_0),
\\
\bar{C}_{22}(t)=\bar{C}_{22}+\phantom{}_{0}\Delta\bar{C}_{22}\, \phantom{}_{0}\xi_{2}(t-t_0),
\\
\bar{S}_{22}(t)=\bar{S}_{22}+\phantom{}_{0}\Delta\bar{S}_{22}\, \phantom{}_{0}\xi_{2}(t-t_0),
\end{array}
\right.
\end{equation}
with the oscillation function
\begin{equation}
\phantom{}_{0}\xi_{2}(t-t_0)=1-\phantom{}_0\tau_2(t)\cos\left(2\pi\phantom{}_{0}f_{2}(t-t_0)\right)
\end{equation}
and the attenuation
\begin{equation}
\phantom{}_0\tau_2(t)=\exp\left(-\frac{2\pi\phantom{}_{0}f_{2}(t-t_0)}{2\phantom{}_{0}Q_{2}}\right).
\end{equation}
The forms of $\mathcal{CA}$ and $\mathcal{DA}_1$ are identical to the ones for $N=2$ static gravity field. The term $\mathcal{DA}_2$ is negligible compared to $\mathcal{DA}_1$, and hence omitted here. The term $\mathcal{DA}_3$ reads 
\begin{equation}
\begin{split}
&\mathcal{DA}_3=C_{00}\,\phantom{}_0\tau_2(t)\left[\cos\left(2\pi f_{00}^+t+\zeta_{00}^+\right)+\cos\left(2\pi f_{00}^-t+\zeta_{00}^-\right)\right]
\\
&+C_{02}\,\phantom{}_0\tau_2(t)\left[\cos\left(2\pi f_{02}^+t+\zeta_{02}^+\right)+\cos\left(2\pi f_{02}^-t+\zeta_{02}^-\right)\right]
\\
&+C_{10}\,\phantom{}_0\tau_2(t)\left[\cos\left(2\pi f_{10}^+t+\zeta_{10}^+\right)+\cos\left(2\pi f_{10}^-t+\zeta_{10}^-\right)\right]
\\
&+C_{20}\,\phantom{}_0\tau_2(t)\left[\cos\left(2\pi f_{20}^+t+\zeta_{20}^+\right)+\cos\left(2\pi f_{20}^-t+\zeta_{20}^-\right)\right]
\\
&+C_{1-2}\,\phantom{}_0\tau_2(t)\left[\cos\left(2\pi f_{1-2}^+t+\zeta_{1-2}^+\right)+\cos\left(2\pi f_{1-2}^-t+\zeta_{1-2}^-\right)\right]
\\
&+C_{12}\,\phantom{}_0\tau_2(t)\left[\cos\left(2\pi f_{12}^+t+\zeta_{12}^+\right)+\cos\left(2\pi f_{12}^-t+\zeta_{12}^-\right)\right]
\\
&+C_{2-2}\,\phantom{}_0\tau_2(t)\left[\cos\left(2\pi f_{2-2}^+t+\zeta_{2-2}^+\right)+\cos\left(2\pi f_{2-2}^-t+\zeta_{2-2}^-\right)\right]
\\
&+C_{22}\,\phantom{}_0\tau_2(t)\left[\cos\left(2\pi f_{22}^+t+\zeta_{22}^+\right)+\cos\left(2\pi f_{22}^-t+\zeta_{22}^-\right)\right].
\end{split}
\end{equation}
It is a summation of several oscillation functions with the attenuation $\phantom{}_0\tau_2(t)$. The frequency, amplitude, and phase of each term are listed below in their order of appearance in the equation. 
\newline
(1) Terms with the frequencies $\left\{f_{00}^{+},f_{00}^{-}\right\}$:
\begin{equation}
f_{00}^+=f_{00}^-=\phantom{}_0f_2,
\end{equation}
\begin{equation}
C_{00}=\frac{3\sqrt{5}}{4}\frac{GM}{R^2}\left(\frac{R}{a}\right)^4\left(3(\sin i)^2-2\right)\sin \left(\frac{\gamma}{2}\right)\phantom{}_0\Delta\bar{C}_{20},
\end{equation}
\begin{equation}
\zeta_{00}^+=\zeta_{00}^-=-2\pi\phantom{}_0f_2t_0.
\end{equation}
(2) Terms with the frequencies $\left\{f_{02}^{+},f_{02}^{-}\right\}$:
\begin{equation}
\left\{
\begin{array}{c}
f_{02}^+=\phantom{}_0f_2-2f_o,
\\
f_{02}^-=\phantom{}_0f_2+2f_o,
\end{array}
\right.
\end{equation}
\begin{equation}
\begin{split}
&C_{02}=-\frac{3\sqrt{5}}{4}\frac{GM}{R^2}\left(\frac{R}{a}\right)^4(\sin i)^2
\\
&\times\left(2\sin \gamma \cos \left(\frac{\gamma}{2}\right)+3\sin \left(\frac{\gamma}{2}\right)\cos \gamma\right)\phantom{}_0\Delta\bar{C}_{20},
\end{split}
\end{equation}
\begin{equation}
\left\{
\begin{array}{c}
\zeta_{02}^+=-2\pi\phantom{}_0f_2t_0-2\omega_{o0},
\\
\zeta_{02}^-=-2\pi\phantom{}_0f_2t_0+2\omega_{o0}.
\end{array}
\right.
\end{equation}
(3) Terms with the frequencies $\left\{f_{10}^{+},f_{10}^{-}\right\}$:
\begin{equation}
\left\{
\begin{array}{c}
f_{10}^+=\phantom{}_0f_2+f_e,
\\
f_{10}^-=\phantom{}_0f_2-f_e,
\end{array}
\right.
\end{equation}
\begin{equation}
C_{10}=\frac{1}{2}\sqrt{\left(\phantom{}_0T_{210}^c\right)^2+\left(\phantom{}_0T_{210}^s\right)^2},
\end{equation}
\begin{equation}
\left\{
\begin{array}{c}
\phantom{}_0T_{210}^c=+3\sqrt{15}\frac{GM}{R^2}\left(\frac{R}{a}\right)^4\sin i\cos i\sin \left(\frac{\gamma}{2}\right)\phantom{}_0\Delta\bar{S}_{21},
\\
\phantom{}_0T_{210}^s=-3\sqrt{15}\frac{GM}{R^2}\left(\frac{R}{a}\right)^4\sin i\cos i\sin \left(\frac{\gamma}{2}\right)\phantom{}_0\Delta\bar{C}_{21},
\end{array}
\right.    
\end{equation}
\begin{equation}
\left\{
\begin{array}{c}
\zeta_{10}^+=-2\pi\phantom{}_0f_2t_0-\omega_{e0}-\phantom{}_0\gamma_{210},
\\
\zeta_{10}^-=-2\pi\phantom{}_0f_2t_0+\omega_{e0}+\phantom{}_0\gamma_{210},
\end{array}
\right.
\end{equation}
\begin{equation}
\tan\phantom{}_0\gamma_{210}=-\phantom{}_0T_{210}^s/\phantom{}_0T_{210}^c.
\end{equation}
(4) Terms with the frequencies $\left\{f_{20}^{+},f_{20}^{-}\right\}$:
\begin{equation}
\left\{
\begin{array}{c}
f_{20}^+=\phantom{}_0f_2+2f_e,
\\
f_{20}^-=\phantom{}_0f_2-2f_e,
\end{array}
\right.
\end{equation}
\begin{equation}
C_{20}=\frac{1}{2}\sqrt{\left(\phantom{}_0T_{220}^c\right)^2+\left(\phantom{}_0T_{220}^s\right)^2},
\end{equation}
\begin{equation}
\left\{
\begin{array}{c}
\begin{split}
\phantom{}_0T_{220}^c=&-\frac{3\sqrt{15}}{2}\frac{GM}{R^2}\left(\frac{R}{a}\right)^4
\\
&\times\left((\cos i)^2-1\right)\sin \left(\frac{\gamma}{2}\right)\phantom{}_0\Delta\bar{C}_{22},
\end{split}
\\
\begin{split}
\phantom{}_0T_{220}^s=&-\frac{3\sqrt{15}}{2}\frac{GM}{R^2}\left(\frac{R}{a}\right)^4
\\
&\times\left((\cos i)^2-1\right)\sin \left(\frac{\gamma}{2}\right)\phantom{}_0\Delta\bar{S}_{22},
\end{split}
\end{array}
\right.    
\end{equation}
\begin{equation}
\left\{
\begin{array}{c}
\zeta_{20}^+=-2\pi\phantom{}_0f_2t_0-2\omega_{e0}-\phantom{}_0\gamma_{220},
\\
\zeta_{20}^-=-2\pi\phantom{}_0f_2t_0+2\omega_{e0}+\phantom{}_0\gamma_{220},
\end{array}
\right.
\end{equation}
\begin{equation}
\tan\phantom{}_0\gamma_{220}=-\phantom{}_0T_{220}^s/\phantom{}_0T_{220}^c.
\end{equation}
(5) Terms with the frequencies $\left\{f_{1\mp2}^{+},f_{1\mp2}^{-}\right\}$:
\begin{equation}
\left\{
\begin{array}{c}
f_{1\mp2}^+=\phantom{}_0f_2+f_e\pm2f_o,
\\
f_{1\mp2}^-=\phantom{}_0f_2-f_e\mp2f_o,
\end{array}
\right.
\end{equation}
\begin{equation}
C_{1\mp2}=\frac{1}{2}\sqrt{\left(\phantom{}_0T_{21\mp2}^c\right)^2+\left(\phantom{}_0T_{21\mp2}^s\right)^2},
\end{equation}
\begin{equation}
\left\{\begin{array}{c}
\begin{split}
&\phantom{}_0T_{21\mp2}^c=-\frac{\sqrt{15}}{2}\frac{GM}{R^2}\left(\frac{R}{a}\right)^4\sin i\left(\cos i\mp1\right) \\
&\times\left(2\sin \gamma \cos \left(\frac{\gamma}{2}\right)+3\sin \left(\frac{\gamma}{2}\right)\cos \gamma\right)\phantom{}_0\Delta\bar{S}_{21},
\end{split}
\\
\begin{split}
&\phantom{}_0T_{21\mp2}^s=+\frac{\sqrt{15}}{2}\frac{GM}{R^2}\left(\frac{R}{a}\right)^4\sin i\left(\cos i\mp1\right) \\
&\times\left(2\sin \gamma \cos \left(\frac{\gamma}{2}\right)+3\sin \left(\frac{\gamma}{2}\right)\cos \gamma\right)\phantom{}_0\Delta\bar{C}_{21},
\end{split}
\end{array}\right.
\end{equation}
\begin{equation}
\left\{
\begin{array}{c}
\zeta_{1\mp2}^+=-2\pi\phantom{}_0f_2t_0-\omega_{e0}\pm2\omega_{o0}-\phantom{}_0\gamma_{21\mp2},
\\
\zeta_{1\mp2}^-=-2\pi\phantom{}_0f_2t_0+\omega_{e0}\mp2\omega_{o0}+\phantom{}_0\gamma_{21\mp2},
\end{array}
\right.
\end{equation}
\begin{equation}
\tan\phantom{}_0\gamma_{21\mp2}=-\phantom{}_0T_{21\mp2}^s/\phantom{}_0T_{21\mp2}^c.
\end{equation}
(6) Terms with the frequencies $\left\{f_{2\mp2}^{+},f_{2\mp2}^{-}\right\}$:
\begin{equation}
\left\{
\begin{array}{c}
f_{2\mp2}^+=\phantom{}_0f_2+2f_e\pm2f_o,
\\
f_{2\mp2}^-=\phantom{}_0f_2-2f_e\mp2f_o,
\end{array}
\right.
\end{equation}
\begin{equation}
C_{2\mp2}=\frac{1}{2}\sqrt{\left(\phantom{}_0T_{22\mp2}^c\right)^2+\left(\phantom{}_0T_{22\mp2}^s\right)^2},
\end{equation}
\begin{equation}
\left\{\begin{array}{c}
\begin{split}
&\phantom{}_0T_{22\mp2}^c=-\frac{\sqrt{15}}{4}\frac{GM}{R^2}\left(\frac{R}{a}\right)^4\left(\cos i\mp1\right)^2 \\ 
&\times\left(2\sin (\gamma)\cos (\frac{\gamma}{2})+3\sin \left(\frac{\gamma}{2}\right)\cos (\gamma)\right)\phantom{}_0\Delta\bar{C}_{22},
\end{split}
\\
\begin{split}
&\phantom{}_0T_{22\mp2}^s=-\frac{\sqrt{15}}{4}\frac{GM}{R^2}\left(\frac{R}{a}\right)^4\left(\cos i\mp1\right)^2 \\ 
&\times\left(2\sin (\gamma)\cos (\frac{\gamma}{2})+3\sin \left(\frac{\gamma}{2}\right)\cos (\gamma)\right)\phantom{}_0\Delta\bar{S}_{22},
\end{split}
\end{array}\right.
\end{equation}
\begin{equation}
\left\{
\begin{array}{c}
\zeta_{2\mp2}^+=-2\pi\phantom{}_0f_2t_0-2\omega_{e0}\pm2\omega_{o0}-\phantom{}_0\gamma_{22\mp2},
\\
\zeta_{2\mp2}^-=-2\pi\phantom{}_0f_2t_0+2\omega_{e0}\mp2\omega_{o0}+\phantom{}_0\gamma_{22\mp2},
\end{array}
\right.
\end{equation}
\begin{equation}
\tan\phantom{}_0\gamma_{22\mp2}=-\phantom{}_0T_{22\mp2}^s/\phantom{}_0T_{22\mp2}^c.
\end{equation}
The resulting frequencies are integer linear combinations of the $_0S_2$ mode frequency $_0f_2$, the Earth's rotation frequency $f_e$, and the satellites' orbital frequency $f_o$, i.e., $_0f_2$, $_0f_2\mp 2f_o$, $_0f_2\pm f_e$, $_0f_2\pm 2f_e$, $_0f_2\pm \left(f_e\mp 2f_o\right)$, $_0f_2\pm \left(2f_e\mp 2f_o\right)$. The amplitudes are functions of the coefficients $\phantom{}_{0}\Delta\bar{C}_{20}$, $\phantom{}_{0}\Delta\bar{C}_{21}$, $\phantom{}_{0}\Delta\bar{S}_{21}$, $\phantom{}_{0}\Delta\bar{C}_{22}$, $\phantom{}_{0}\Delta\bar{S}_{22}$ and TianQin's orbital parameters $a$, $i$, $\gamma$. The phases are determined by $_0f_2$, the earthquake occurrence time $t_0$, the initial phases of the Earth's rotation and the satellite orbits, $\omega_{e0}$ and $\omega_{o0}$, as well as the parameters $a$, $i$, $\gamma$. The frequencies and associated amplitudes correspond to the peaks of the ASD curves in the presence of the free oscillations.





\bibliography{apssamp}

\end{document}